\documentclass[11pt]{article}

\usepackage[cp1251]{inputenc}
\usepackage[T1]{fontenc}
\usepackage{textcomp}
\usepackage[centertags]{amsmath}
\usepackage{amsfonts}
\usepackage{amssymb}
\usepackage{revsymb}
\usepackage[pdftex]{hyperref}
\usepackage{graphicx}
\usepackage{graphbox}
\usepackage[numbers,sort&compress]{natbib}

\usepackage{paperinitial}

\paperinitialization{15mm}{15mm}{15mm}{15mm}{2pt}{10pt}

\DeclareMathOperator{\sgn}{sgn}

\newcommand{\lan}{\langle}
\newcommand{\ran}{\rangle}

\newcommand{\e}{\varepsilon}
\newcommand{\vf}{\varphi}

\newcommand{\s}{\sigma}

\newcommand{\al}{\alpha}
\newcommand{\be}{\beta}
\newcommand{\ga}{\gamma}

\newcommand{\de}{\delta}

\newcommand{\la}{\lambda}

\newcommand{\vs}{\varsigma}

\newcommand{\spx}{\mathbf{x}}

\newcommand{\spe}{\mathbf{e}}

\begin{document}
\allowdisplaybreaks[4]
\frenchspacing
\setlength{\unitlength}{1pt}

\title{{\Large\textbf{Generation of twisted photons by\\ undulators filled with dispersive medium}}}

\date{}

\author{O.V. Bogdanov${}^{1),2)}$\thanks{E-mail: \texttt{bov@tpu.ru}},\; P.O. Kazinski${}^{1)}$\thanks{E-mail: \texttt{kpo@phys.tsu.ru}},\; and G.Yu. Lazarenko${}^{1),2)}$\thanks{E-mail: \texttt{laz@phys.tsu.ru}}\\[0.5em]
{\normalsize ${}^{1)}$ Physics Faculty, Tomsk State University, Tomsk 634050, Russia}\\
{\normalsize ${}^{2)}$ Division for Mathematics and Computer Sciences},\\
{\normalsize Tomsk Polytechnic University, Tomsk 634050, Russia}}

\maketitle

\begin{abstract}

The radiation of twisted photons by undulators filled with a homogeneous dielectric dispersive medium is considered. The general formulas for the average number of radiated twisted photons are obtained. The radiation of undulators in the dipole regime and the radiation of the helical and planar wigglers are studied in detail. It is shown that the selection rules for radiation of twisted photons established for undulators in a vacuum also holds for undulators filled with a dielectric medium. In the case of a medium with plasma permittivity the lower undulator harmonics do not form. This fact can be used for generation of twisted photons with nonzero orbital angular momentum on the lowest admissible harmonic. The use of the effect of inverse radiation polarization for generation of twisted photons with larger  orbital angular momentum is described. The influence of the anomalous Doppler effect on the projection of the total angular momentum of radiated twisted photons is investigated. The parameters of the undulator and the charged particles are found such that the produced radiation, in particular, the Vavilov-Cherenkov radiation, is a pure source of twisted photons with definite nonzero orbital angular momentum. The developed theory is used to describe the radiation of twisted photons by beams of electrons and protons in the undulators filled with helium. We also consider the radiation of X-ray twisted photons by electrons in the undulator filled with xenon. The parameters are chosen so as to be achievable at the present experimental facilities.


\end{abstract}

\section{Introduction}

The undulator radiation represents a unique source of photons that combines a high degree of coherence and intensity with a large flexibility of its parameters and availability at the acceleration facilities. Nowadays the free-electron lasers (FELs) employing the undulator radiation are the standard tool for generation of an intense flux of coherent photons from THz up to X-ray spectral ranges \cite{NovoFEL,XFEL,HemStuXiZh14,Hemsing7516,Rubic17,PRRibic19,Gover19rmp}. Usually the chamber where the charged particles move in the undulator or the whole undulator are mounted in an ultra high vacuum. Nevertheless, there are theoretical and experimental works where the undulators and FELs filled with a dielectric medium are studied \cite{GinzbThPhAstr,BazZhev77,GevKorkh,BarysFran,SahKot05,SahKotGrig,KotSah12,KonstKonst,Appolonov,GrichSad,Reid93,Pantell90,YarFried,Reid89prl,ReidPant,Pantell86,Fisher88,Pantell89,Feinstein89prl,ArutOgan94}. The presence of a medium in the undulator degrades the properties of the electron beam evolving in it but, on the other hand, one may adjust the parameters of the beam and the medium in such a way that the undulator will produce the photons with larger energies and narrower spectral bands than the vacuum undulator for a given energy of electrons. The degradation of the electron beam can be overcome. This is possible even in FELs where the coherence properties of radiation strongly depend on the beam configuration \cite{Reid93,Pantell90,Reid89prl,ReidPant,YarFried,Fisher88,Pantell89,Feinstein89prl}. The use of proton beams makes this problem virtually negligible though, of course, the generation of undulator radiation becomes a much harder task.

The undulator radiation is known to be a source of twisted photons \cite{SasMcNu,AfanMikh,BordKN,HemMar12,BHKMSS,HKDXMHR,RibGauNin14,KatohPRL,KatohSRexp,BKL2,ABKT,EpJaZo,BKb,BKL4,EppGusel19,BKL6,BKLb,Rubic17,HemStuXiZh14}, i.e., the source of the quanta of the electromagnetic field with the definite energy, the momentum projection onto the direction of propagation, the projection of the total angular momentum onto this axis, and the helicity \cite{GottfYan,JaurHac,BiaBirBiaBir,JenSerprl,JenSerepj,PadgOAM25,Roadmap16,KnyzSerb,TorTorTw,AndBabAML,TorTorCar,MolTerTorTor,MolTerTorTorPRL}. In the present paper, we investigate the influence of the homogeneous dispersive dielectric medium loaded in the undulator on the properties of radiation of twisted photons. In \cite{BKL5} the general theory was developed for the radiation of twisted photons by charged particles moving in an inhomogeneous dispersive medium. We apply this theory to describe the radiation of twisted photons by undulators filled with a medium. As a result, some general properties and selection rules for the radiation produced by these undulators are established. In particular, we find that the selection rules for radiation of twisted photons in vacuum undulators are also fulfilled for radiation of twisted photons in undulators filled with a homogeneous dispersive medium.

It turns out that the use of these undulators offers certain advantages in generation of twisted photons in comparison with the vacuum undulators in addition to the merits mentioned above. Namely, the radiation of lower harmonics generated in undulators filled with a dielectric medium having a plasma permittivity do not form for sufficiently large plasma frequencies. This allows one to generate the photons possessing a nonzero orbital angular momentum at the lowest radiation harmonic. In the case of vacuum undulators, this is possible only by the use of helically microbunched beams of particles \cite{HKDXMHR,HemStuXiZh14}. Furthermore, suitably adjusting the energy of charged particles and the parameters of the medium, one can twist the Vavilov-Cherenkov (VC) radiation produced in the helical undulator. The photons of this radiation possessing zero projection of the total angular momentum become almost completely circularly polarized. Then the projection of their orbital angular momentum is $-s$, where $s$ is the helicity of radiated photons. The use of planar wigglers filled with a dielectric medium allows one to obtain the VC radiation with nonzero projection of the total angular momentum. However, the probability of radiation of twisted photons obeys the reflection symmetry \cite{BKL3,BKL6,BKL2} in this case.

Another interesting effect that we investigate in the paper is the inverse polarization of the helical undulator radiation in the paraxial regime. This effect manifests itself as domination of the radiation polarization that is inverse to the chirality of helical trajectory of a charged particle in the undulator. This effect also exists in the vacuum helical undulators (see, e.g., \cite{Bord.1,BKL4}) but, for certain parameters, it becomes more pronounced in the helical undulators filled with a medium. For a given positive harmonic of undulator radiation, the twisted photons with parameters belonging to the domain of inverse polarization possess the modulus of the projection of the orbital angular momentum by $2\hbar$ more than in the domain of the usual radiation polarization. The negative harmonics of undulator radiation appear when the generation of VC radiation becomes possible. These harmonics correspond to the anomalous Doppler effect \cite{GrichSad,GinzbThPhAstr,BazZhev77,BarysFran,Nezlin,KuzRukh08,ShiNat18}. We study the influence of the anomalous Doppler effect on the properties of twisted photons produced in undulators. The twisted photons radiated at these harmonics in the helical undulator carry the projection of the total angular momentum $m=\vs n$, where $\vs=\pm1$ is the chirality of the particle trajectory. In other words, the selection rule for radiation of twisted photons in helical undulators \cite{EppGusel19,BKL4,BKL2,KatohSRexp,KatohPRL,BHKMSS,SasMcNu} remains intact in the presence of a homogeneous dielectric medium but the sign of the harmonic number $n$ may take both values.

As examples, we consider the radiation of electrons and protons in undulators and wigglers filled with helium. We also describe the radiation of X-ray twisted photons with the projection of the orbital angular momentum $l=2$ produced by electrons in the undulator filled with xenon. In the latter case, the radiation is created near the photoabsorption $M$-edge of xenon. The same mechanism for  generation of X-ray VC radiation from the beam of protons traversing the target made of amorphous carbon was experimentally verified in \cite{BazylevXrayVC}. The X-ray VC radiation from other materials was also observed \cite{BazylZhev,KvdWLV}. The use of X-ray VC radiation for diagnostics of particle beams can be found in \cite{XrayVC2,XrayVC1}. Notice that the angular momentum of radiated twisted photons in all the considered examples can be shifted to larger values by employing the addition rule valid for the radiation produced by helically microbunched beams of charged particles \cite{HemMar12,HKDXMHR,TrHemsing,ExpHemsing,BKLb}. The intensity of radiation can be increased with the help of the coherent radiation created by a periodic train of bunches, the frequency of one of the coherent harmonics must coincide with the energy of twisted photons radiated by one charged particle \cite{Ginzburg,KuzRukh08,Gover19rmp,PRRibic19,Rubic17,Hemsing7516,HemStuXiZh14,KKST,SPTS}. Currently, the highest number of a distinguishable coherent harmonic of the electron bunch train is of order $100$ with the corresponding energy of photons $474$ eV \cite{PRRibic19}.

The paper is organized as follows. In Sec. \ref{Gener_Form_Sec}, the general formulas for the average number of twisted photons produced by charged particles traversing a dielectric plate are presented. We find the transformation law of the mode functions of twisted photons and establish the selection rules for their radiation. The formula for the average number of twisted photons created by Gaussian and helically microbunched beams of identical charged particles is also given. In Sec. \ref{Undul_Sec}, we derive the average number of twisted photons radiated in the undulator filled with a homogeneous dispersive medium. We start with the estimates of multiple scattering of the radiating charged particles and find the restrictions on the parameters of the particle beam and the medium when this scattering can be neglected in describing the properties of radiation. Then, in Sec. \ref{Dip_Appr_Sec}, we obtain the general formula for the average number of twisted photons radiated by the undulator in the dipole regime. The properties of the energy spectrum of radiated photons are also discussed. Section \ref{Hel_Wig_Sec} is devoted to the helical wiggler filled with a medium. We derive the formula for the average number of twisted photons produced by it and analyze the polarization properties of this radiation paying a special attention to the influence of the radiation polarization on the orbital angular momentum of radiated twisted photons. In Sec. \ref{Plan_Wigg_Sec}, we completely describe the radiation of a planar wiggler filled with a medium in terms of the twisted photons. In Conclusion we summarize the results. Throughout the paper we use the system of units such that $\hbar=c=1$ and $e^2=4\pi\al$, where $\al\approx1/137$ is the fine structure constant.

\section{General formulas}\label{Gener_Form_Sec}

In the paper \cite{BKL5}, the formalism was developed that allows one to describe the radiation of twisted photons by charged particles moving in an inhomogeneous dispersive medium. In particular, the formula was obtained for the probability to detect a twisted photon created by the charged particle passing through the dielectric plate of the width $L$ possessing the permittivity $\e(k_0)>0$ (see Sec. V.A of \cite{BKL5}). For the reader convenience, we present here some general formulas from that paper. Notice that such a dielectric plate can be a gas or liquid confined into the cuvette of a proper form.

Let the twisted photon recorded by the detector possess the helicity $s$, the projection of the total angular momentum $m$ to the axis $3$ (the axis $z$), the projection of the momentum $k_3$, and the modulus of the perpendicular component of the momentum $k_\perp$. The energy of such a photon is
\begin{equation}
    k_0=\sqrt{k_\perp^2+k_3^2},
\end{equation}
and its state in the Coulomb gauge in the vacuum is characterized by the wave function \cite{GottfYan,JaurHac,BiaBirBiaBir,JenSerprl,JenSerepj,KnyzSerb,BKL2}
\begin{equation}\label{tw_phot_vac}
\begin{gathered}
    \psi_3(m,k_3,k_\perp;\spx)=j_m(k_\perp x_+,k_\perp x_-) e^{ik_3x_3},\qquad \psi_\pm(s,m,k_3,k_\perp;\spx)=\frac{in_\perp}{s\pm n_3}\psi_3(m\pm1,k_3,k_\perp;\spx),\\
    \boldsymbol{\psi}(s,m,k_3,k_\perp;\spx)= \frac12\big[\psi_-(s,m,k_3,k_\perp;\spx)\spe_+ +\psi_+(s,m,k_3,k_\perp;\spx)\spe_-
    \big]+\psi_3(m,k_3,k_\perp;\spx)\spe_3,
\end{gathered}
\end{equation}
where $n_\perp:=k_\perp/k_0$, $n_3:=k_3/k_0$, and $\spe_\pm=\spe_1\pm i\spe_2$. The basis unit vectors $\{\spe_1,\spe_2,\spe_3\}$ constitute a right-handed triple and the unit vector $\spe_3$ is directed along the $z$ axis.

The presence of the dielectric plate changes the vacuum mode functions of twisted photons. Let the dielectric plate be situated at $z\in[-L,0]$ and the detector of twisted photons be located in the vacuum in the region $z>0$. We assume that the typical size of the plate along the $x$ and $y$ axes is large and neglect the influence of the edge effects on the properties of radiation \cite{Pafomov}. As a rule, this is justified for the radiation of relativistic particles as long as their radiation is concentrated in a narrow cone. In the presence of the dielectric plate, the mode functions of the twisted photons have the form
\begin{equation}\label{vac_mode_f}
\begin{alignedat}{2}
    a&\boldsymbol{\psi}(s,m,k_3,k_\perp)&\quad&\text{for $z>0$},\\
    a&\big[b_+\boldsymbol{\psi}'(1,m,k'_3,k_\perp) +b_-\boldsymbol{\psi}'(-1,m,k'_3,k_\perp) +(k'_3\leftrightarrow-k'_3)\big]&\quad&\text{for $z\in[-L,0]$},\\
    a&\big[a_+\boldsymbol{\psi}(1,m,k_3,k_\perp) +d_+\boldsymbol{\psi}(1,m,-k_3,k_\perp) +a_-\boldsymbol{\psi}(-1,m,k_3,k_\perp) +d_-\boldsymbol{\psi}(-1,m,-k_3,k_\perp)\big]&\quad&\text{for $z<-L$},
\end{alignedat}
\end{equation}
where
\begin{equation}
    \psi'_3(m,k'_3,k_\perp)=\psi_3(m,k'_3,k_\perp),\qquad\psi'_\pm(s',m,k'_3,k_\perp)=\frac{in_\perp}{s'\e^{1/2}(k_0)\pm n'_3}\psi'_3(m\pm1,k'_3,k_\perp),
\end{equation}
and
\begin{equation}\label{mass_shell_plate2}
    n'_3:=k_3'/k_0=\sqrt{\e(k_0)-n_\perp^2}.
\end{equation}
Also
\begin{equation}\label{Fresnel_coeff}
\begin{split}
    b_\pm&=\frac{\e^{1/2}\pm s}{4\e n'_3}(\pm sn'_3+\e^{1/2}n_3),\\
    a_{\pm}&=\frac{2 (1\pm s)\e n_3n'_3\cos(k'_3L)-i(\e^{2}n_{3}^{2}+n'^2_{3}\pm s\e(n_{3}^{2}+n'^2_{3}))\sin(k'_3L)}{4\e n_3n'_3}e^{ik_3L}, \\
    d_{\pm}&=-i\frac{\e^{2}n_{3}^{2}-n'^2_{3}\pm s\e(n_{3}^{2}-n'^2_{3})}{4\e n_3n'_3}\sin(k'_3L) e^{-ik_3L},
\end{split}
\end{equation}
where $s$ is the helicity of the mode function \eqref{vac_mode_f}. The coefficients \eqref{Fresnel_coeff} obey the unitarity relation
\begin{equation}
    1+|d_+|^2+|d_-|^2=|a_+|^2+|a_-|^2,
\end{equation}
for real-valued $k_3$. The constant $a$ is found from the normalization of the mode functions:
\begin{equation}
    |a|^{-2}=|a_+|^2+|a_-|^2=\Big|1+\frac18\Big[(\e^2+1)\Big(\frac{n^2_3}{n'^2_3} +\frac{n'^2_3}{\e^2 n^2_3}\Big)-4\Big]\sin^2(k'_3 L)\Big|.
\end{equation}
The above formulas can be generalized to the case of the medium with absorption \cite{BKL5}.

Let us denote as $\boldsymbol{\Phi}(s,m,k_\perp,k_3;\spx)$ the mode function \eqref{vac_mode_f}. Then the probability to record the twisted photon created by the particles with charges $e_l$ is given by
\begin{multline}\label{prob_plate}
    dP(s,m,k_\perp,k_3)=|a|^2\bigg|\sum_{l}e_l\int_{-\infty}^\infty d\tau e^{-ik_0 x^0_l(\tau_l)}
    \Big\{\dot{x}_{3l}(\tau_l)\Phi_3(s,m,k_\perp,k_3;\spx_l(\tau_l))+\\
    +\frac12\big[\dot{x}_{+l}(\tau_l)\Phi_-(s,m,k_\perp,k_3;\spx_l(\tau_l)) +\dot{x}_{-l}(\tau_l)\Phi_+(s,m,k_\perp,k_3;\spx_l(\tau_l))\big] \Big\} \bigg|^2 n_\perp^3\frac{dk_3 dk_\perp}{16\pi^2},
\end{multline}
where $x_l^\mu(\tau)$ are the world lines of particles. Strictly speaking, the quantity \eqref{prob_plate} is the probability to detect a twisted photon only in the first Born approximation with respect to the classical current of the charged particle. If one neglects the quantum recoil and replaces the current operator by a c-number quantity, then the equations of quantum electrodynamics are exactly solvable. In that case, the probability to record a twisted photon is expressed though \eqref{prob_plate}, whereas the expression \eqref{prob_plate} is equal to the average number of radiated twisted photons (see for details \cite{BKL2,BKL4,BKL5,ippccb}).

Notice some general properties of the expression \eqref{prob_plate}. Let $A(s,m,k_\perp,k_3;j]$ be the amplitude of radiation of a twisted photon by the current $j_i$ entering into \eqref{prob_plate}. Then on rotating the current $j_i(x)$ around the detector axis by the angle of $\vf$, $j_i\rightarrow j_i^\vf$, the amplitude transforms as
\begin{equation}
    A(s,m,k_\perp,k_3;j^\vf]=e^{im\vf} A(s,m,k_\perp,k_3;j],
\end{equation}
i.e., it has the same transformation law as in a vacuum. Consequently, the selection rules established in Sec. 2 of \cite{BKL3} also hold for the radiation of twisted photons by charged particles in the presence of a dielectric plate. Furthermore, the radiation produced by the current of particles moving parallel to some plane containing the detector axis, i.e., the current being such that $\arg j_+=const$, obeys the reflection symmetry
\begin{equation}\label{refl_symm}
    dP(s,m,k_\perp,k_3)=dP(-s,-m,k_\perp,k_3).
\end{equation}
The proof of this relation is the same as it was given in \cite{BKL2} for the radiation from a charged particle moving along a planar trajectory in a vacuum. Selecting suitably the axes $x$ and $y$, one may put $j_+(x)=j_-(x)$. Then, performing the rotation around the $z$ axis by the angle of $\pi$, we obtain that $j^\pi_+(x)=j^\pi_-(x)$ and
\begin{equation}
    j_3\rightarrow j^\pi_3,\qquad j_\pm\rightarrow -j^\pi_\pm,\qquad\Phi_3(s,m)\rightarrow\Phi_3(-s,-m),\qquad \Phi_\pm(s,m)\rightarrow-\Phi_\mp(-s,-m).
\end{equation}
Whence
\begin{equation}
    A(s,m,k_\perp,k_3;j_\pi]=A(-s,-m,k_\perp,k_3;j]=e^{im\pi}A(s,m,k_\perp,k_3;j].
\end{equation}
Taking the modulus squared of the both parts of the last equality, we arrive at the relation \eqref{refl_symm}.

Further, we shall need the formula for the probability of radiation of twisted photons by the beam of identical charged particles with small dispersion of the initial momenta (see for more detail \cite{BKb,BKLb}). This formula is deduced from the general formula \eqref{prob_plate}. The radiation produced by a helically microbunched beam of particles with the helix pitch $\de$ and the chirality $\chi_b=\pm1$ is concentrated at the harmonics
\begin{equation}\label{coherent_harm}
    k_0=2\pi\chi_b n_c\be_3/\de,\qquad \chi_b n_c>0,\;n_c\in \mathbb{Z},
\end{equation}
where $\be_3$ is the velocity of particles along the axis $3$. At these harmonics, the average number of radiated twisted photons becomes
\begin{equation}\label{dP_bunch}
    dP_\rho(s,m,k_\perp,k_3)=N\sum_{j=-\infty}^\infty f_{m-j}dP_1(s,j,k_\perp,k_3) +N(N-1)|\bar{\vf}_{n_c}|^2dP_1(s,m-n_c,k_\perp,k_3),
\end{equation}
where $N$ is the number of particles, $dP_1(s,m,k_\perp,k_3)$ is the average number of twisted photons created by one charged particle moving long the center of the beam, $f_{m}$ and $\bar{\vf}_{n}$ are the incoherent and coherent interference factors, respectively. The explicit expressions for these factors are presented in \cite{BKb,BKLb}. If $k_3\de\gg1$ and $k_3\s_3\gg1$, where $\s_3$ is the longitudinal dimension of the beam, then the coherent contribution is strongly suppressed and formula \eqref{dP_bunch} is valid for any energy of the radiated photon. Notice that the usual Gaussian beam of particles is obtained from the helically microbunched one when $\de\gg\s_3$.

\section{Undulator}\label{Undul_Sec}

Consider the radiation of twisted photons by one charged particle moving in the undulator filled with the dielectric medium with permittivity $\e(k_0)$. The trajectory of the particle has the form (see, e.g., \cite{Bord.1})
\begin{equation}\label{traj_undul}
    x^i(t)=r^i(t)+\be^i t,\qquad \be^i=(0,0,\be_3),\qquad t\in[-TN_u,0],\qquad L=TN_u,
\end{equation}
where $t$ is the laboratory time, $\be_3\in[0,1)$, $r^i(t)$ is a periodic function of $t$ with the period $T=:2\pi/\omega$ and zero average over this period, $N_u\gg1$ is the number of undulator sections, the length of one section is $\la_0=2\pi\be_3/\omega$. For $t<-TN_u$ or $t>0$, the particle moves along the undulator axis with the velocity
\begin{equation}
    \be_\parallel:=\sqrt{1-1/\ga^2}.
\end{equation}
The parts of the particle trajectory are joined continuously at the instants $t=-TN_u$ and $t=0$. The undulator strength parameter $K$ is determined by the relation
\begin{equation}
    \be_3^2=1-\frac{1+K^2}{\ga^2}.
\end{equation}
It can also be defined as
\begin{equation}
    K^2=\ga^2\lan\be_\perp^2\ran,
\end{equation}
where the angular brackets denote the average over the trajectory period. The magnetic field strength in the undulator is
\begin{equation}
    eH=\frac{\omega m_p}{z\be_3} \sqrt{2}K,
\end{equation}
where $m_p$ is the mass of a charged particle and $z$ is its charge in the units of the elementary charge. In the relativistic case, $\ga\gg1$, the following estimates for the parameters of particle's trajectory hold
\begin{equation}\label{rel_appr}
    r^2_{1,2}\sim\frac{K^2}{\omega^2\ga^2},\qquad |r_3|\lesssim\frac{K^2}{2\pi\omega\ga^2},\qquad \frac{K}{\ga}\ll1,\qquad\be_3\approx1-\frac{1+K^2}{2\ga^2}.
\end{equation}
We also assume throughout this paper that the axis of the detector of twisted photons coincides with the undulator axis. Then the periodic part of the particle trajectory \eqref{traj_undul} is written as
\begin{equation}
    \textbf{r}=\frac12(r_+\spe_- +r_-\spe_+)+r^3\spe_3,
\end{equation}
where $r_\pm=r^1\pm ir^2$.

We will be interested in the contribution to radiation of twisted photons produced by the charged particle on the part of the trajectory $t\in[-TN_u,0]$, i.e., we will discard the contribution of transition radiation. Such an approximation is valid for sufficiently large $N_u$. Besides, we will assume that the multiple scattering of the charged particle on the particles of the medium does not considerably affect the properties of radiation produced by one particle. This is justified (see, e.g., \cite{BazylevXrayVC,BazylZhev}) when the average square of the multiple scattering angle \cite{BazylZhev,BazylevXrayVC,Migdal57,TMikaelian,PDG10},
\begin{equation}
    q\approx\Big(\frac{zE_s}{m_p\ga\be_\parallel^2}\Big)^2\frac{L}{L_{rad}}=\frac{4\pi z^2}{\al\ga^2\be_\parallel^4}\frac{m_e^2L}{m_p^2L_{rad}},
\end{equation}
satisfies
\begin{equation}
    q\lesssim n_\perp^2,\qquad q\lesssim \lan\be_\perp^2\ran=K^2/\ga^2.
\end{equation}
Here $m_e$ is the electron mass and $L_{rad}$ is the radiation length. The formula for the radiation length in the medium consisting of single type nuclei with the charge $Z$ reads as \cite{PDG10}
\begin{equation}
    L_{rad}^{-1}=\frac{M_p}{716.4}n_m Z(Z+1)\ln\frac{287}{Z^{1/2}},
\end{equation}
where $M_p$ is the proton mass in grams and $n_m$ is the particle number of the medium in $1$ cm${}^{3}$. To run such an undulator in the FEL regime, more stringent conditions on the parameters of the particle beam and the dielectric medium must be imposed \cite{Reid93,Pantell90,Reid89prl,ReidPant,YarFried,Fisher88,Pantell89,Feinstein89prl,Appolonov}.

\subsection{Dipole approximation}\label{Dip_Appr_Sec}

Let us find, at first, the average number of twisted photons radiated by a charged particle in the undulator in the dipole regime when
\begin{equation}\label{dipole_appr}
    |k'_3r_3|\ll1,\qquad k_\perp|r_+|\ll1,\qquad K\ll1.
\end{equation}
The evaluation of the radiation amplitude entering into \eqref{prob_plate} is reduced to the evaluation of the integrals
\begin{equation}\label{I_integrals}
\begin{split}
    I_3&=\int_{-TN_u}^0 dt\dot{x}_3e^{-ik_0t+ik_3'x_3(t)}j_m\big(k_\perp x_+(t),k_\perp x_-(t)\big),\\
    I_\pm&=\frac{in_\perp}{s'\e^{1/2}\mp n_3'}\int_{-TN_u}^0 dt\dot{x}_\pm e^{-ik_0t+ik_3'x_3(t)}j_{m\mp1}\big(k_\perp x_+(t),k_\perp x_-(t)\big).
\end{split}
\end{equation}
In the dipole approximation \eqref{dipole_appr}, we have
\begin{equation}\label{j_m_expans}
\begin{split}
    j_m&=\de_{m0}\Big(1-\frac14k_\perp^2|r_+|^2\Big) +\frac12\de_{m1}k_\perp r_+ -\frac12\de_{m,-1}k_\perp r_-+\cdots,\\
    j_{m+1}&=\de_{m,-1} +\frac12\de_{m0}k_\perp r_+ -\frac12\de_{m,-2}k_\perp r_-+\cdots,\\
    j_{m-1}&=\de_{m1} +\frac12\de_{m2}k_\perp r_+ -\frac12\de_{m0}k_\perp r_-+\cdots,\\
\end{split}
\end{equation}
where the properties of the functions $j_m$ have been used (see (A3), (A4) of \cite{BKL2}). For brevity, we do not write out the arguments of the functions $j_m$.

\begin{figure}[tp]
\centering
\includegraphics*[width=0.49\linewidth]{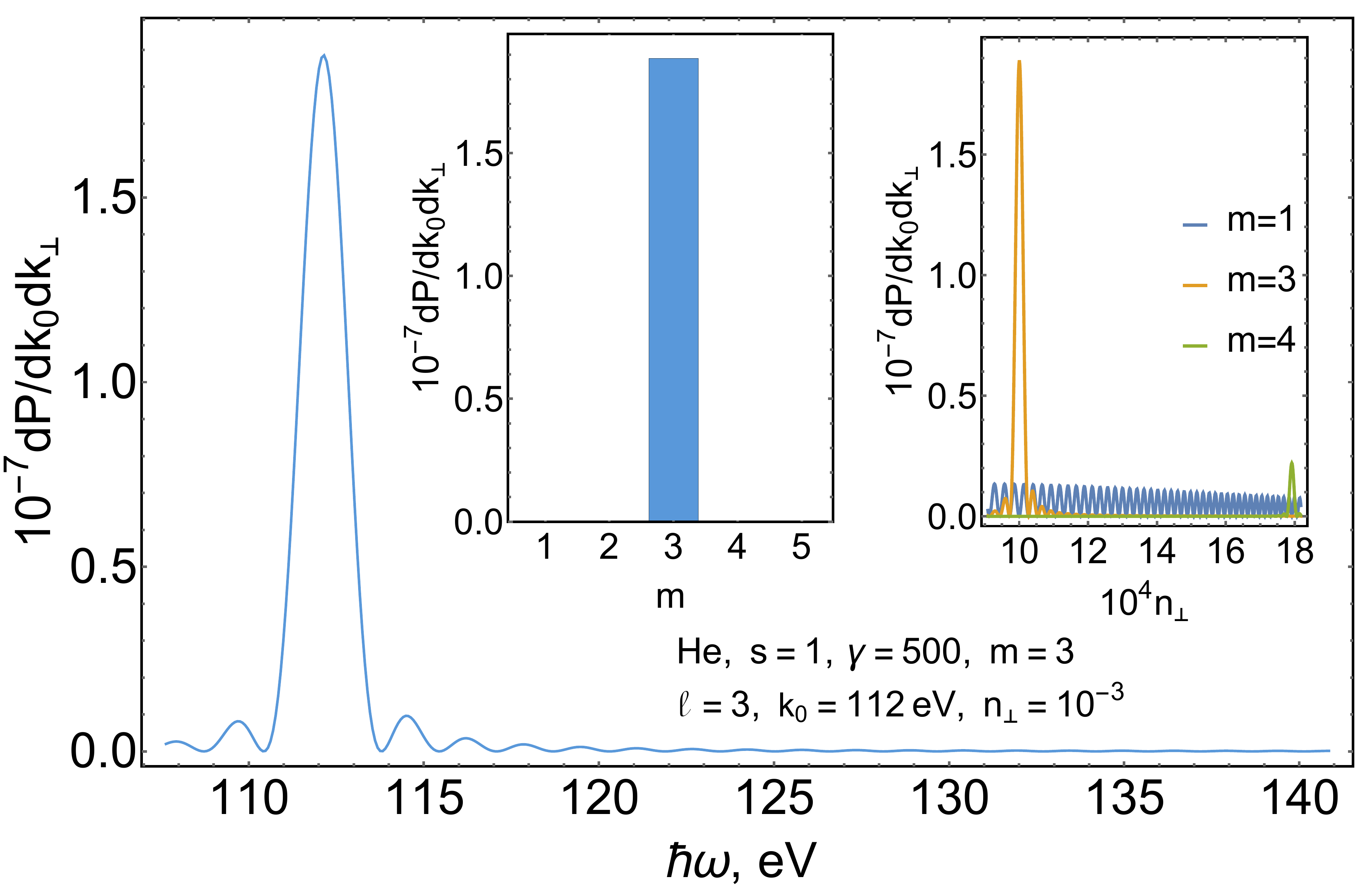}\,
\includegraphics*[width=0.49\linewidth]{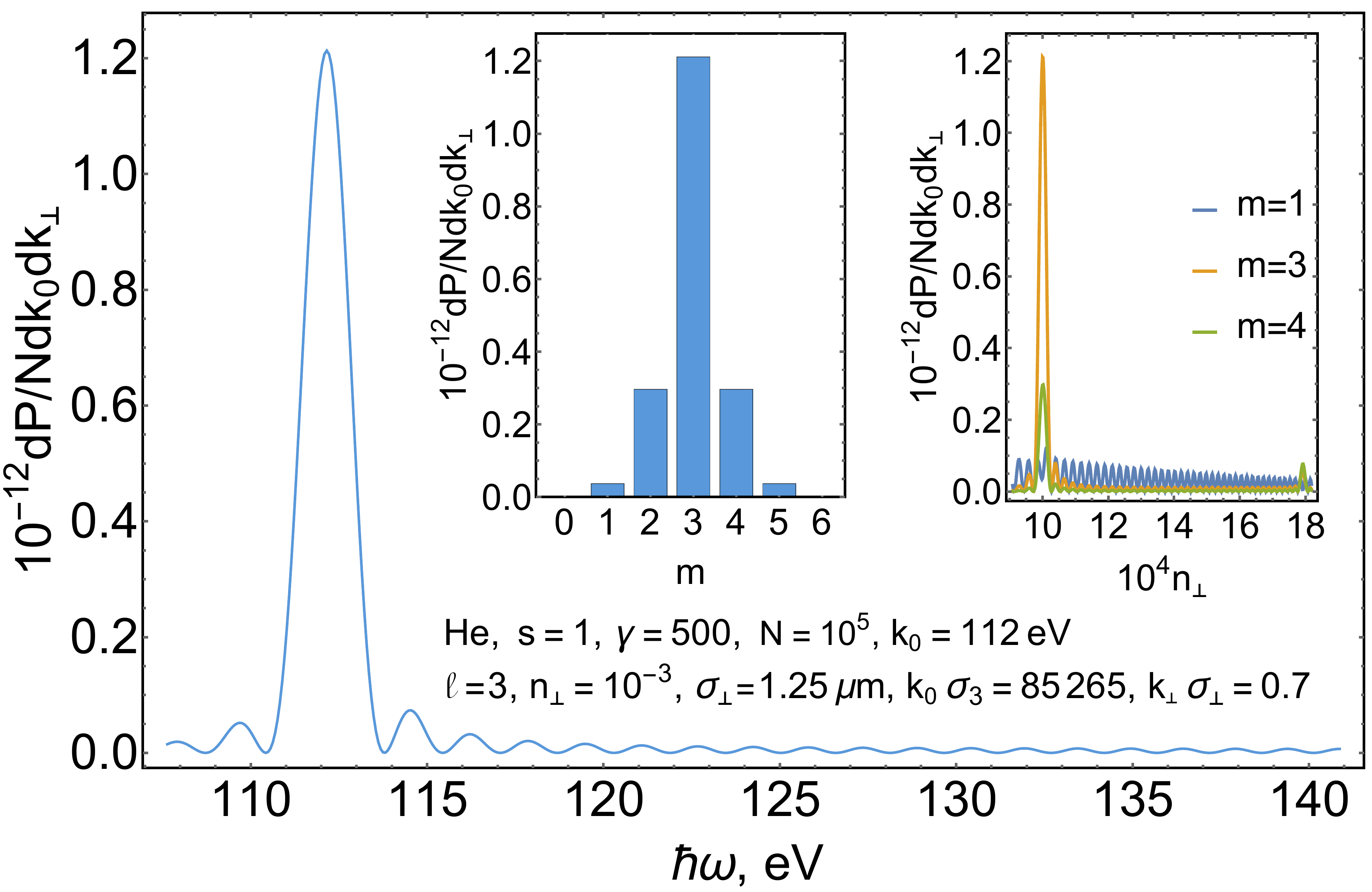}
\caption{{\footnotesize The average number of twisted photons produced by electrons moving in the helical undulator with the chirality $\vs=1$. The Lorentz factor is $\ga=500$, $E=256$ MeV. The undulator is filled with helium under the pressure $1/4$ atm and the temperature $0$ ${}^o$C. The gas concentration is calculated as $n_m=p/(k_BT)$. In accordance with formula \eqref{plasm_freq}, the plasma frequency $\omega_p\approx0.14$ eV. The number of undulator periods $N_u=40$ and the period $\la_0=1$ cm. The undulator strength parameter $K=1/5$, which corresponds to the magnetic field strength $H=3.03$ kG in the undulator. The projection of the total angular momentum of radiated photons per photon is $\ell=3$ and the energy of photons at the third harmonic is $k_0=112$ eV. The ratios $n_\perp^2/q\approx7.3$ and $K^2/(q\ga^2)\approx1.2$ that means that the multiple scattering can be neglected. The dependence $k_0(n_\perp)$ for the different harmonics is presented on the left panel in Fig. \ref{znk_plots}. At the given energy of photons, the harmonics with $n<3$ are not formed (see the plot for $n=m=1$). On the left panel: The radiation from one electron is described. The radiation of twisted photons with $s=-1$ is strongly suppressed. On the right panel: The radiation from the beam of electrons is considered. The beam is supposed to have a Gaussian profile with the longitudinal dimension $\s_3=150$ $\mu$m (duration $0.5$ ps) and the transverse size $\s_\perp=1.25$ $\mu$m. The coherent contribution to radiation of twisted photons is strongly suppressed. We see that the radiation of photons with projection of the orbital angular momentum $l=m-s=2$ dominates.}}
\label{plsm_perm_plots}
\end{figure}

It is useful to represent the periodic part of the trajectory as a Fourier series
\begin{equation}\label{traj_Four}
    \mathbf{r}=\sum_{n=-\infty}^\infty \mathbf{r}_n e^{i\omega n t},\qquad \mathbf{r}_0=0.
\end{equation}
Substituting the expansions \eqref{j_m_expans}, \eqref{traj_Four} into \eqref{I_integrals} and neglecting $k_3'r_3$ in the exponent, we obtain
\begin{equation}
\begin{split}
    I_3&=\de_{m0}\vf_0\Big(\beta_3 -\frac14 k_\perp^2\sum_{n=-\infty}^\infty|r_{n+}|^2\Big) +\frac12 \de_{m1}\sum_{n=-\infty}^\infty k_\perp r_{n+}\vf_n -\frac12 \de_{m,-1}\sum_{n=-\infty}^\infty k_\perp r_{n-}\vf_n,\\
    I_\pm&=\frac{n_\perp}{s'\e^{1/2}\mp n'_3}\Big[\de_{m0}\vf_0\sum_{n=-\infty}^\infty \frac{k_\perp\omega n}{2}|r_{n+}|^2 -\de_{m,\pm1}\sum_{n=-\infty}^\infty \omega nr_{n\pm}\vf_n \Big],
\end{split}
\end{equation}
where the terms giving a negligible contribution to the radiation probability are discarded (see the estimates \eqref{rel_appr}, \eqref{dipole_appr}) and
\begin{equation}\label{vfn}
    \vf_n:=2\pi e^{iT N_u[k_0(1-n_3'\be_3)-n\omega]/2}\de_{N_u}\big(k_0(1-n_3'\be_3)-n\omega\big),\qquad \de_{N_u}(x):=\frac{\sin(T N_u x/2)}{\pi x}.
\end{equation}
Then the contribution to the radiation amplitude of a twisted photon that comes from the wave function $\psi'(s',m,k'_3,k_\perp)$ is written as
\begin{multline}
    I_3+\frac12(I_++I_-)=\de_{m0}\vf_0\Big[\beta_3 -\frac14\sum_{n=-\infty}^\infty(k_\perp^2-2s'\e^{1/2}k_0\omega n)|r_{n+}|^2 \Big]+\\
    +\frac12\de_{m1}\sum_{n=-\infty}^\infty\Big[1 -\frac{\omega n}{k_0(s'\e^{1/2}-n'_3)}\Big] k_\perp r_{n+}\vf_n -\frac12\de_{m,-1}\sum_{n=-\infty}^\infty \Big[1 +\frac{\omega n}{k_0(s'\e^{1/2}+n'_3)}\Big] k_\perp r_{n-}\vf_n.
\end{multline}
As a result, the average number of twisted photons produced by one relativistic charged particle in the undulator in the dipole regime becomes
\begin{equation}\label{dP_dip}
    dP(s,m,k_\perp,k_3)=|zea|^2\big(\de_{m0}|B_0|^2 +\de_{m1}|B_+|^2 +\de_{m,-1}|B_{-}|^2\big) n_\perp^3\frac{dk_3dk_\perp}{64\pi^2},
\end{equation}
where
\begin{equation}
\begin{split}
    B_0&=\vf_0\Big[\Big(\frac1{\e}+\frac{n_3}{n_3'}\Big)\Big(\beta_3 -\frac14\sum_{n=-\infty}^\infty k_\perp^2|r_{n+}|^2\Big) +\frac{s}{2} \Big(1+\frac{n_3}{n'_3}\Big) \sum_{n=-\infty}^\infty k_0\omega n|r_{n+}|^2 \Big],\\
    B_\pm&=\sum_{n=-\infty}^\infty \frac{k_\perp r_{n\pm}}{2}\Big\{\frac{1}{\e}+\frac{n_3}{n'_3} -\frac{\omega n}{k_\perp n_\perp}\Big[ n_3 +\frac{n_3'}{\e} \pm s\Big(1+\frac{n_3}{n'_3}\Big)\Big] \Big\}\vf_n+(k'_3\leftrightarrow-k'_3).
\end{split}
\end{equation}
For $N_u\gtrsim10$, the functions $|\vf_n(\s k_3')|$, $\s=\pm1$, possess sharp maxima at
\begin{equation}\label{harmonics}
    k_0=\frac{n\omega}{1-\s n'_3\be_3}\;\Leftrightarrow\;n_\perp=\sqrt{\e(k_0)-(1-n\omega/k_0)^2/\be_3^2},
\end{equation}
where $n_\perp\in[0,1]$. These relations determine the undulator radiation spectrum. The peculiarities of this spectrum will be discussed below in Sec. \ref{spectrum_sect}. The plots of the average number of twisted photons produced in the helical undulator filled with helium are presented in Fig. \ref{plsm_perm_plots}.

Neglecting the terms that are small for $N_u\gtrsim10$, we have
\begin{equation}\label{Bpm}
    |B_\pm|^2=\sideset{}{'}\sum_{n=-\infty}^\infty \frac{k_\perp^2 |r_{n\pm}|^2}{8}\Big|\frac{1}{\e}+\frac{n_3}{n'_3} -\frac{\omega n}{k_\perp n_\perp}\Big[ n_3 +\frac{n_3'}{\e} \pm s\Big(1+\frac{n_3}{n'_3}\Big)\Big] \Big|^2|\vf_n|^2+(k'_3\leftrightarrow-k'_3),
\end{equation}
where the prime at the sum sign reminds that only those $n$ are kept that satisfy \eqref{n_interv}. It is supposed that there are unforbidden harmonics from the intervals \eqref{n_interv} contributing considerably to \eqref{dP_dip} for a given $k_0$. As we see, in the dipole regime, the most part of undulator radiation consists of the twisted photons with the projection of the total angular momentum $m=\{-1,0,1\}$. Recall that, in the dipole regime, the undulator in a vacuum radiates mainly the twisted photons with $m=\pm1$ \cite{BKL2}. The contribution with $m=0$ in the case of the undulator filled with a medium corresponds to the VC radiation. Also note that in the ultraparaxial regime when
\begin{equation}\label{ultraparaxial}
    n_\perp^2\ll1,\qquad n_\perp^2\ll\e,\qquad n_\perp^2\ll\e^{1/2}\omega n/k_0,
\end{equation}
the expression \eqref{Bpm} can be simplified to
\begin{equation}
    |B_\pm|^2 \approx \de_{s,\pm1}\big|1+\e^{-1/2}\big|^2 \sideset{}{'}\sum_{n=-\infty}^\infty \frac{\omega^2 n^2 |r_{n\pm}|^2}{2n_\perp^2} |\vf_n|^2+(k'_3\leftrightarrow-k'_3),
\end{equation}
which is valid in the leading order. Thus we see from \eqref{dP_dip} that, in this approximation, the main contribution to the radiation of twisted photons with $m=\pm1$ comes from the twisted photons with the projection of the orbital angular momentum $l=m-s=0$. This value of the orbital angular momentum can be shifted by an integer number by employing the helically microbunched beams of charged particles as a radiator in such an undulator \cite{HKDXMHR,BKLb}.

\subsubsection{Spectrum}\label{spectrum_sect}

Now we discuss some typical characteristics of the radiation spectrum \eqref{harmonics}. For a fixed $k_0$, we have
\begin{equation}\label{n_interv}
    n\in
    \left\{
      \begin{array}{ll}
        \frac{k_0}{\omega}[1-\be_3\e^{1/2},1-\be_3\chi^{1/2}]\cup \frac{k_0}{\omega}[1+\be_3\chi^{1/2},1+\be_3\e^{1/2}], & \hbox{$\chi\geqslant0$;} \\[0.3em]
        \frac{k_0}{\omega}[1-\be_3\e^{1/2},1+\be_3\e^{1/2}], & \hbox{$\chi\in(-1,0)$.}
      \end{array}
    \right.
\end{equation}
where $\chi:=\e-1$. The boundaries $k_0(1\mp\be_3\chi^{1/2})/\omega$ of the intervals of admissible harmonic numbers correspond to the condition of the total internal reflection, $n_\perp=1$. The harmonics with $n>0$ describe the usual undulator radiation modified by the presence of the medium. In this case, $\s=\pm1$. The branch with $\s=-1$ describes the contribution to radiation from the reflected wave and $k_0<n\omega$ in this case, whereas the branch with $\s=1$ comes from the direct wave and $k_0>n\omega$. The harmonics with $n\leqslant0$ are realized only for $\be_3\e^{1/2}\geqslant1$ and $\s=1$. The harmonic $n=0$ corresponds to the VC radiation and \eqref{harmonics} reproduces the condition for the Cherenkov cone with $n_\perp\equiv\sin\theta$ in this case. The harmonics with $n<0$ describe the radiation with the anomalous Doppler effect \cite{GrichSad,GinzbThPhAstr,BazZhev77,BarysFran,Nezlin,KuzRukh08,ShiNat18}. The function $n_\perp(n,k_0)$ is an increasing function of $n$ for $k_0>\omega n$, i.e., for $\s=1$, and a decreasing function of $n$ for $k_0<\omega n$, i.e., for $\s=-1$. Notice also that for $\chi\leqslant0$ and
\begin{equation}\label{harm_forbid}
    k_0(1-\be_3\e^{1/2})>\omega n_0,
\end{equation}
the harmonics with $n\in[1,n_0]$ are not formed. If the inequality \eqref{harm_forbid} is fulfilled for any $k_0$ within the transparence zone of the medium, then the lower harmonics are completely absent.

\begin{figure}[tp]
\centering
\includegraphics*[width=0.48\linewidth]{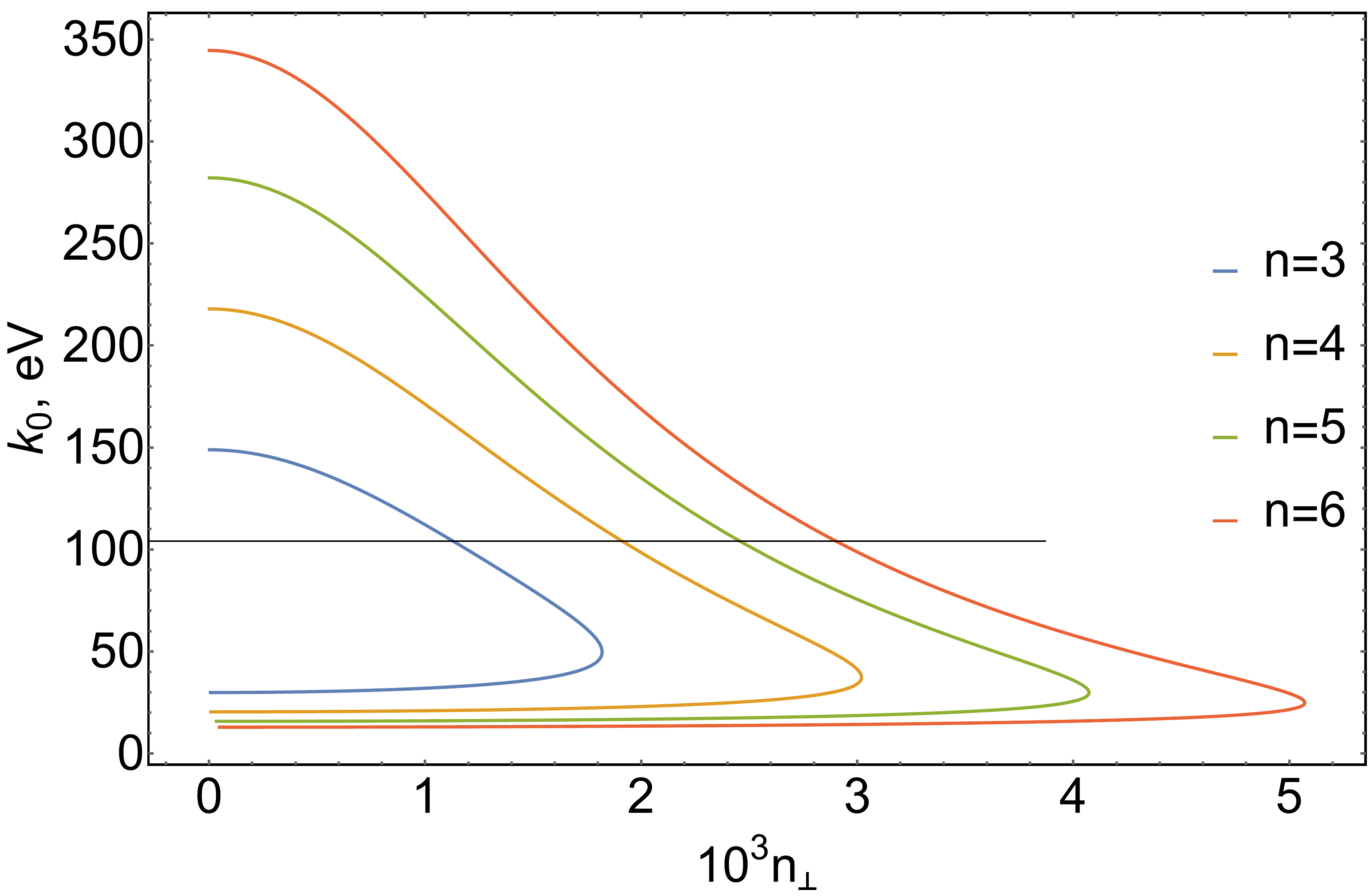}\;
\includegraphics*[width=0.48\linewidth]{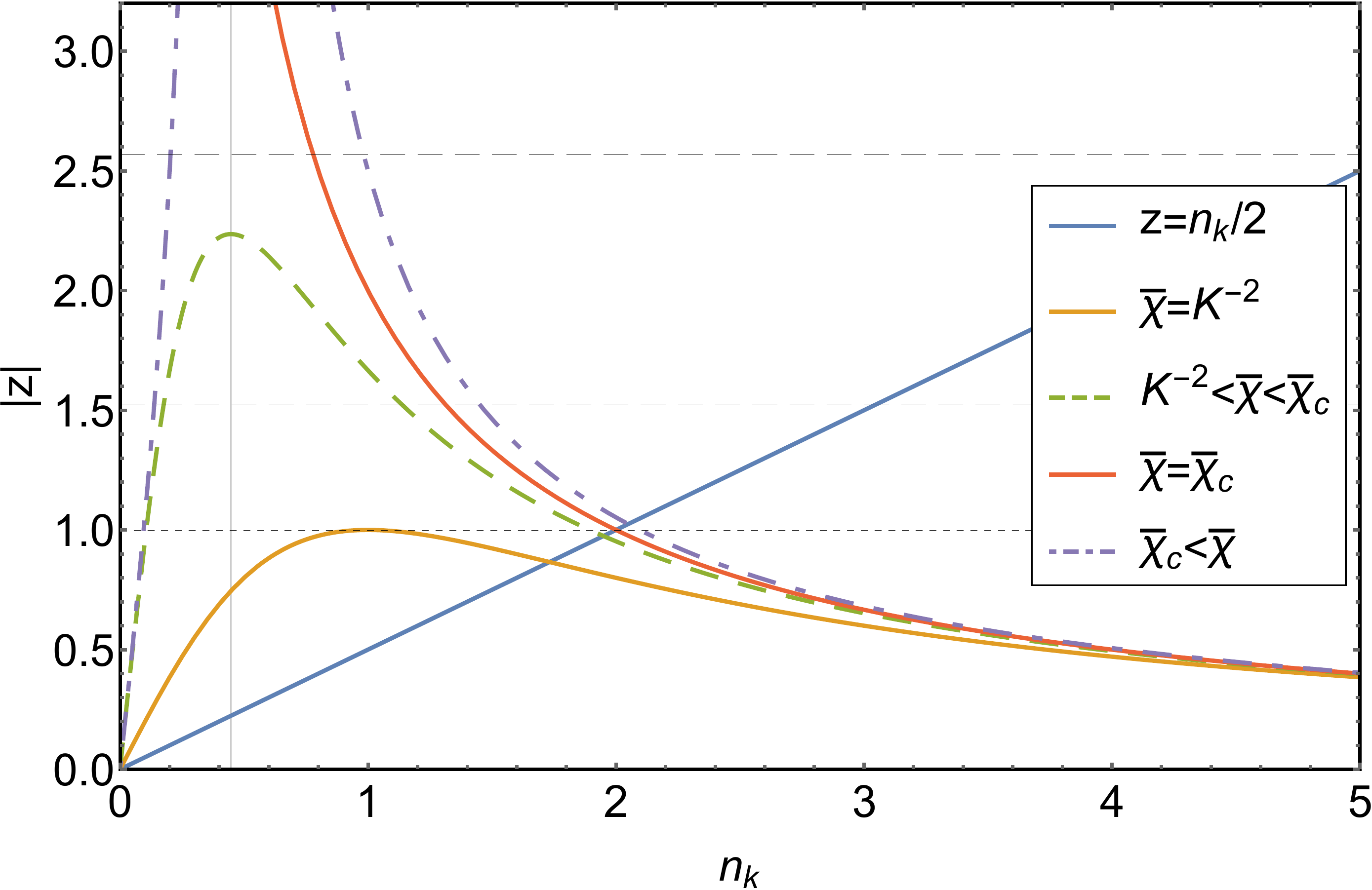}
\caption{{\footnotesize On the left panel: The dependence $k_0(n_\perp)$ for the harmonics $n=\overline{3,6}$ of radiation of twisted photons in the undulator with the parameters described in Fig. \ref{plsm_perm_plots}. The harmonics with $n=\{1,2\}$ are forbidden. The thin horizontal line corresponds to $k_0=112$ eV. On the right panel: The function $|z(n_k)|$ for the different values of $\bar{\chi}$. The undulator strength parameter $K=1/2$. The inclined straight line is $z=\vs\bar{k}_0 n_k/m=n_k/2$. The straight vertical line is $n_k=\sqrt{\bar{\chi}-\bar{\chi}_c}$. The straight horizontal lines: the dotted line is $z=1$; the dashed lines are $z=b_{2,1}\approx1.53$, $z=c_{2,1}\approx2.57$; the thin solid line is $z=b_{1,1}\approx1.84$. The unique intersection point of the line $z=\vs\bar{k}_0 n_k/m$ with the curve $z(n_k)$ gives $n_k$ and, thereby, $\bar{k}_0$ for the radiated twisted photons. Then it is not difficult to see which polarization dominates for these parameters.}}
\label{znk_plots}
\end{figure}

Consider an important particular case of the plasma permittivity in more detail. In this case,
\begin{equation}\label{plasm_perm}
    \e(k_0)=1-\omega^2_p/k_0^2,
\end{equation}
where $\omega_p$ is the plasma frequency. For the material consisting of a single type of nuclei, the following approximate formula holds \cite{LandLifshECM}:
\begin{equation}\label{plasm_freq}
    \omega_p^2=4\pi\al Z n_m/m_e.
\end{equation}
As long as $\chi<0$ in this case, then $n\geqslant1$ and
\begin{equation}
    k_0=\frac{n\omega\pm|\be_3|\sqrt{n^2\omega^2n_3^2-(1-\be_3^2 n_3^2)\omega_p^2}}{1-\be_3^2 n_3^2}.
\end{equation}
The energy of photons radiated at the harmonic $n$ belongs to the interval $k_0\in[k_0^-,k_0^+]$, where
\begin{equation}
    k_0^\pm=\frac{n\omega \pm|\be_3|\sqrt{n^2\omega^2-(1-\be_3^2)\omega_p^2}}{1-\be_3^2}.
\end{equation}
Notice that $k_0^-\geqslant\omega_p$ and $k_0^-=\omega_p$ only for $\omega n=\omega_p$. The maximum value of $n_\perp=n_\perp^c$ at a given harmonic is
\begin{equation}
    (n_\perp^c)^2=\frac{n^2\omega^2+\omega_p^2(\be_3^2-1)}{n^2\omega^2+\omega_p^2\be_3^2}.
\end{equation}
It corresponds to the energy
\begin{equation}
    k_0^c=\frac{n^2\omega^2+\omega_p^2\be_3^2}{n\omega}.
\end{equation}
Besides, if
\begin{equation}\label{prohib_harm}
    n_0<(1-\be_3^2)^{1/2}\omega_p/\omega,
\end{equation}
then the harmonics with $n\in[1,n_0]$ are not formed for any $n_\perp$. This fact can be employed for generation of twisted photons with large projection of the total angular momentum $m$ at the lowest admissible harmonic (see Figs. \ref{plsm_perm_plots} and \ref{znk_plots}). The plots of the dependence $k_0(n_\perp)$ for several lower harmonics are presented in Fig. \ref{znk_plots}.

\subsection{Helical wiggler}\label{Hel_Wig_Sec}

Now we turn to the radiation of undulator in the wiggler regime when the conditions \eqref{dipole_appr} are violated. As in the case of undulator in a vacuum, a simple analytic expression for the average number of twisted photons radiated by a charged particle moving along a trajectory of a general form in the wiggler filled with a medium cannot be derived. Therefore, we consider the two particular cases: the ideal helical wiggler and the planar wiggler.

\begin{figure}[tp]
\centering
\includegraphics*[width=0.47\linewidth]{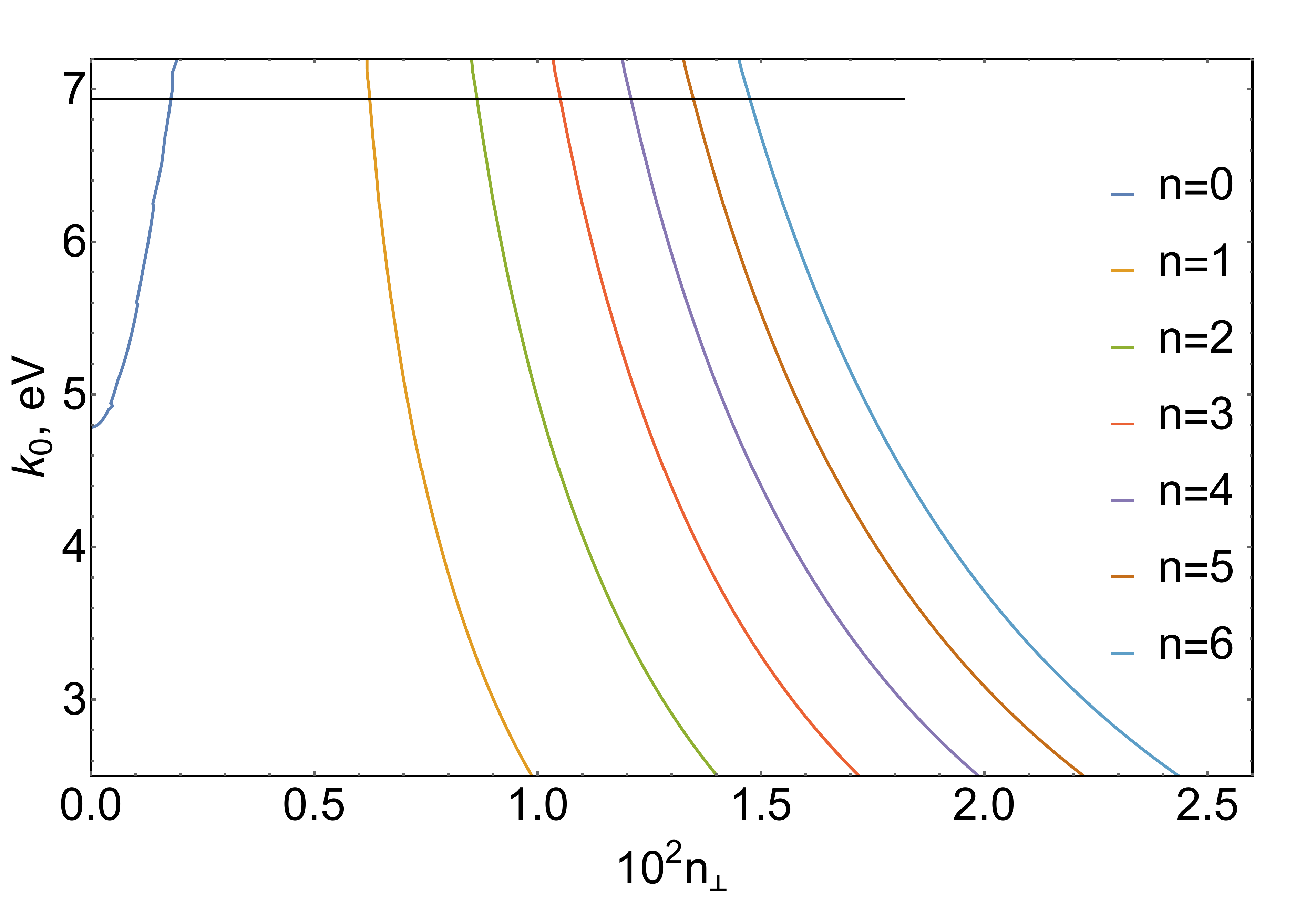}\;
\includegraphics*[width=0.48\linewidth]{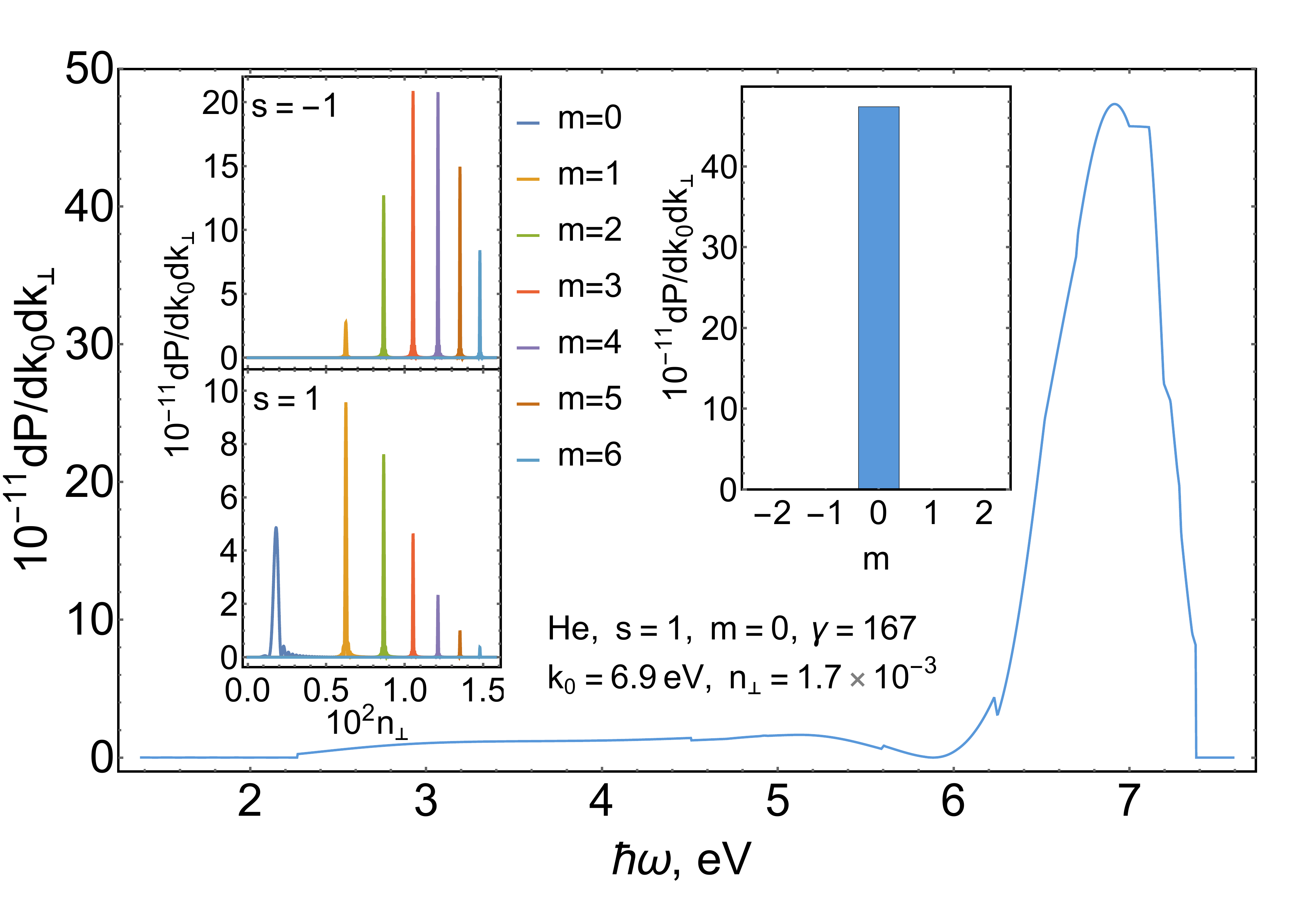}
\caption{{\footnotesize On the left panel: The dependence $k_0(n_\perp)$ for the harmonics $n=\overline{1,6}$ of radiation of twisted photons in the undulator filled with helium under the pressure $1$ atm and the temperature $0$ ${}^o$C. The experimental data for permittivity are taken from \cite{RefrIndex}. On the right panel: The average number of twisted photons produced by one electron moving in such helical wiggler with the chirality $\vs=1$. The Lorentz factor is $\ga=167$, $E=85.3$ MeV. The number of wiggler periods $N_u=25$ and the period $\la_0=1$ cm. The undulator strength parameter $K=1$, which corresponds to the magnetic field strength $H=15.1$ kG in the wiggler. The energy of photons at the zeroth harmonic (the VC radiation) is taken as $k_0=6.9$ eV. The ratios $n_\perp^2/q\approx1.0$ and $K^2/(q\ga^2)\approx12$ that means that the multiple scattering can be neglected. We see that, at the zeroth harmonic, the radiation of photons with $s=1$ is suppressed and so the radiation of photons with projection of the orbital angular momentum $l=-1$ dominates.}}
\label{VC_pol_plots}
\end{figure}

\begin{figure}[tp]
\centering
\includegraphics*[width=0.48\linewidth]{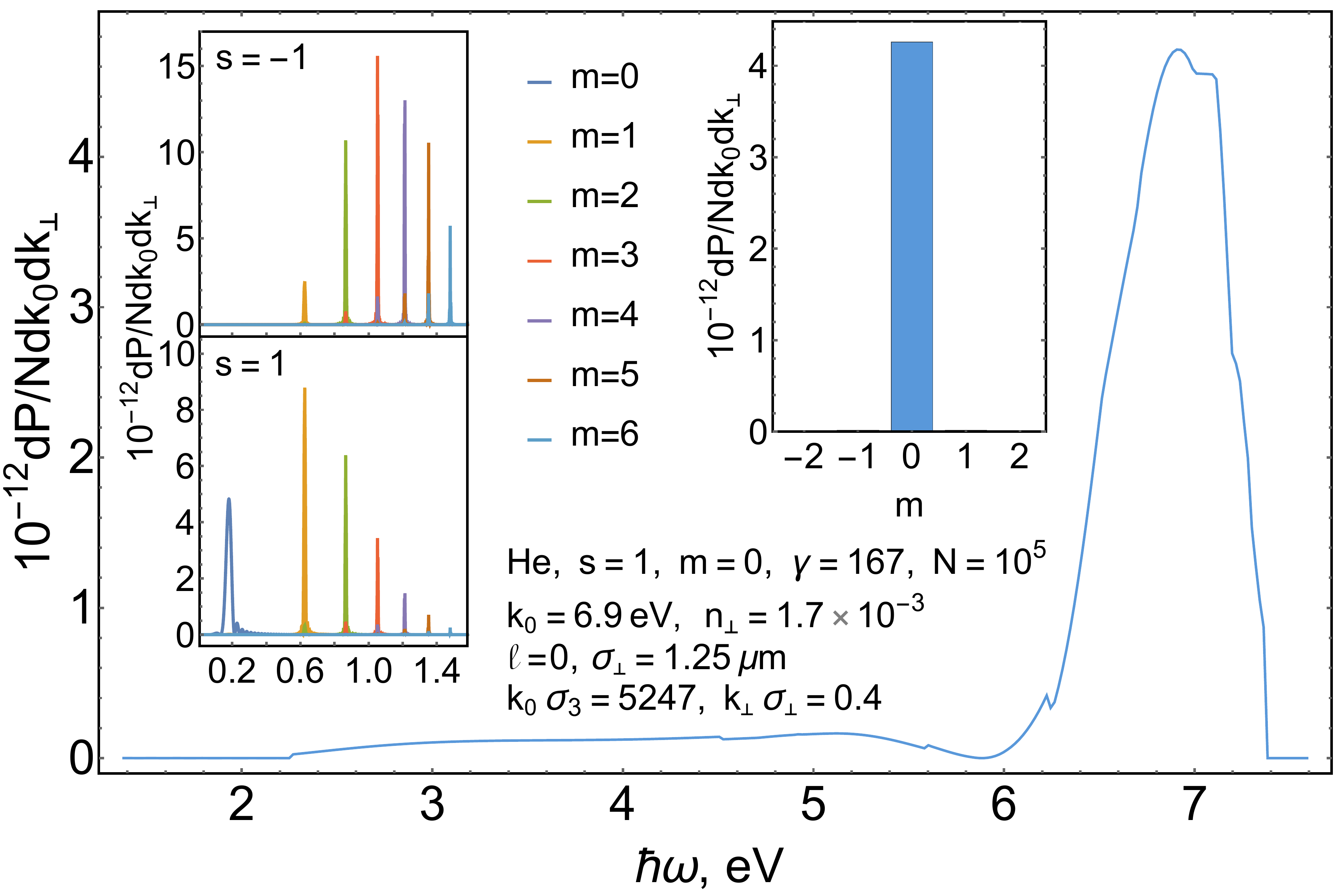}\;
\includegraphics*[width=0.48\linewidth]{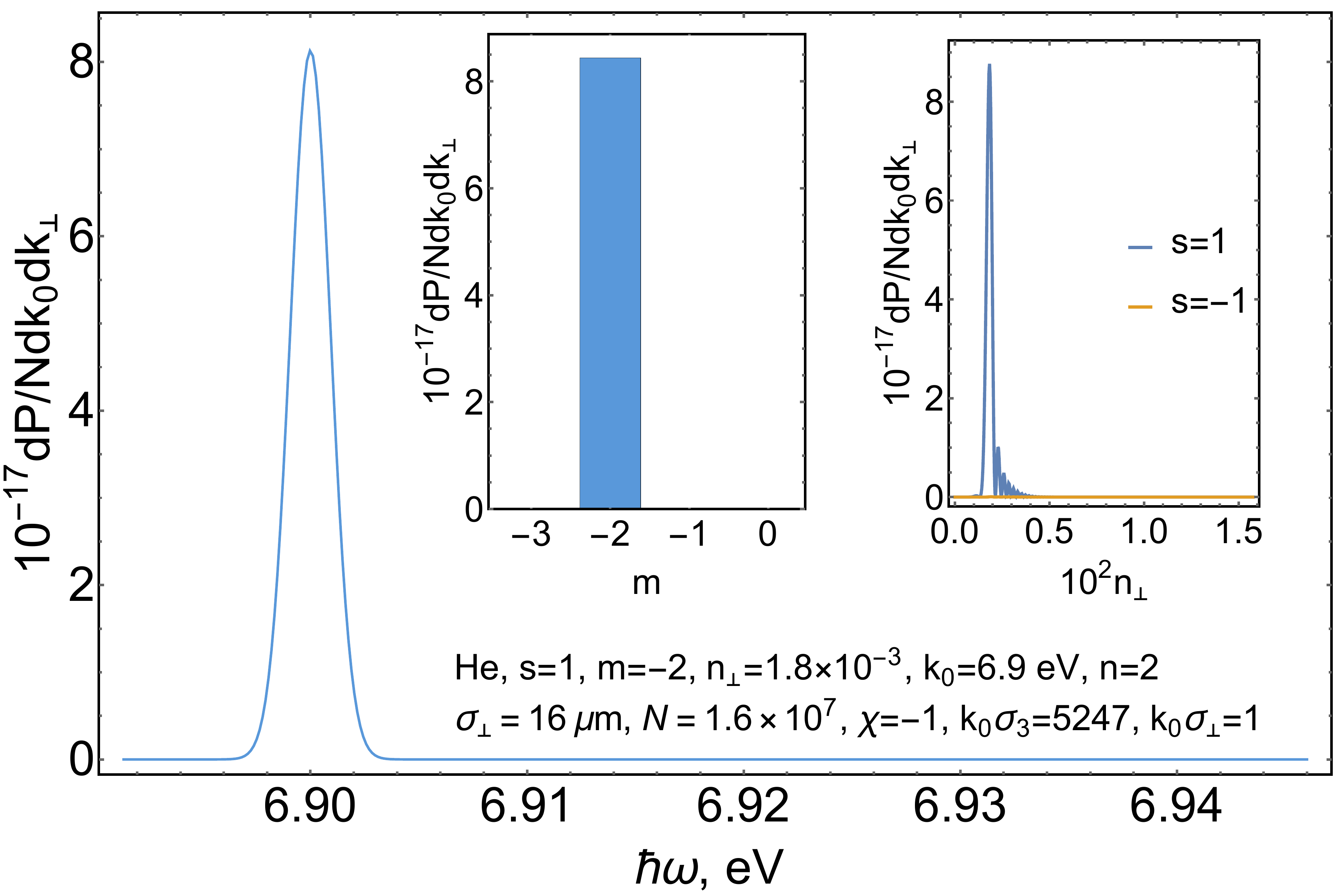}
\caption{{\footnotesize The same as on the left panel in Fig. \ref{VC_pol_plots} but for beams of electrons. On the left panel: The radiation from a Gaussian beam of electrons is considered. We see that, at the zeroth harmonic, the radiation of photons with projection of the orbital angular momentum $l=-1$ dominates. On the right panel: The radiation from a helically microbunched beam of electrons is described. The chirality of the beam $\chi=-1$, the longitudinal dimension $\s_3=150$ $\mu$m (duration $0.5$ ps), and the transverse size $\s_\perp=16$ $\mu$m. The number of coherent harmonic \eqref{coherent_harm} is $n_c=2$ and the helix pitch $\de=0.36$ $\mu$m is chosen such that the coherent radiation is concentrated at $k_0=6.9$ eV. The coherent contribution to radiation dominates and the contribution of higher harmonics is suppressed at the given energy of photons (compare with the plot on the left panel). The fulfillment of the addition rule is clearly seen. The radiation of twisted photons with projection of the orbital angular momentum $l=-3$ prevails.}}
\label{VC_pol_beam_plots}
\end{figure}

We start with the ideal helical wiggler. In this case, the trajectory of the charged particle takes the form \eqref{traj_undul} with
\begin{equation}
    r_\pm=re^{\pm i\vs\omega t},\qquad r_3=0,\qquad K=\ga\omega r,
\end{equation}
where $\omega>0$ and $\vs=\pm1$. Then the integrals \eqref{I_integrals} can easily be evaluated
\begin{equation}
    I_3=\be_3\vf_{\vs m}J_m(k_\perp r),\qquad I_\pm=\mp\vf_{\vs m}\frac{\vs\omega rn_\perp}{\e^{1/2}\mp n'_3}J_{m\mp1}(k_\perp r).
\end{equation}
As a result, neglecting the transition radiation, we arrive at
\begin{equation}
    dP(s,m,k_\perp,k_3)=\bigg|zea\vf_{\vs m}\Big[\Big(\frac{1}{\e} +\frac{n_3}{n'_3}\Big)\Big(\be_3 -\frac{\vs m\omega n_3'}{n_\perp k_\perp}\Big)J_m -\Big(1 +\frac{n_3}{n'_3}\Big)\frac{s\vs K}{n_\perp\ga}J'_m\Big] +(k_3'\leftrightarrow-k_3')\bigg|^2 n_\perp^3\frac{dk_\perp dk_3}{64\pi^2},
\end{equation}
where, for brevity, we omit the arguments of the Bessel functions. For $N_u\gtrsim10$, taking into account \eqref{harmonics}, we have approximately
\begin{equation}\label{dP_hel_wig}
    dP(s,m,k_\perp,k_3)=|zea\vf_{\vs m}|^2\Big|\Big(\frac{1}{\e} +\frac{n_3}{n'_3}\Big)\frac{\e\be_3-n_3'}{n_\perp^2}J_m -\Big(1 +\frac{n_3}{n'_3}\Big)\frac{s\vs K}{n_\perp\ga}J'_m\Big|^2n_\perp^3\frac{dk_\perp dk_3}{64\pi^2} +(k_3'\leftrightarrow-k_3'),
\end{equation}
in the leading order in $N_u$. The radiation spectrum has the form \eqref{harmonics} with $n=\vs m$, i.e., the selection rule is the same as for the ideal helical wiggler in a vacuum. In the relativistic case, the contribution of the reflected wave to \eqref{dP_hel_wig} is strongly suppressed. The plots of the average number of twisted photons produced in the wiggler filled with helium are given in Figs. \ref{VC_pol_plots} and \ref{VC_pol_beam_plots}.

The case when the paraxial approximation holds,
\begin{equation}
    |\chi|\ll1,\qquad n_\perp\ll1,
\end{equation}
is of a peculiar interest. In this case, the projection of the orbital angular momentum of the radiated twisted photon can be introduced as $l:=m-s$. This approximation implies
\begin{equation}\label{spectrum}
    \bar{k}_0=\frac{2\vs m}{K^{-2}+1+n_k^2-\bar{\chi}},\qquad n_k^2=\frac{2\vs m}{\bar{k}_0}+\bar{\chi}-1-K^{-2},\qquad k_\perp r=\bar{k}_0 n_k,
\end{equation}
where $n_k:=n_\perp\ga/K$, $\bar{k}_0:=k_0K^2/(\omega\ga^2)$, and $\bar{\chi}:=\chi\ga^2/K^2$.

\subsubsection{Polarization and angular momentum of radiation}

The degree of polarization of radiation is specified by the ratio
\begin{equation}
    A(s):=\frac{dP(s)}{dP(1)+dP(-1)}.
\end{equation}
In the paraxial approximation and for $m\neq0$,
\begin{equation}\label{polariz}
    A(s)=\frac{\big[J'_m(\vs mz) -s\vs(n_k-1/z)J_m(\vs mz) \big]^2}{2\big[J'^2_m(\vs mz) +(n_k-1/z)^2J^2_m(\vs mz)\big]},\qquad z:=\frac{2n_k}{n_k^2+\bar{\chi}_c-\bar{\chi}}=\frac{\vs\bar{k}_0}{m}n_k,
\end{equation}
where $\bar{\chi}_c:=1+K^{-2}$. For $m=0$, i.e., for the VC radiation, we have
\begin{equation}
    A(s)=\frac{\big[J_1(\bar{k}_0n_k) +s\vs n_kJ_0(\bar{k}_0n_k) \big]^2}{2\big[J^2_1(\bar{k}_0n_k) +n_k^2J^2_0(\bar{k}_0n_k)\big]}.
\end{equation}
All the radiated photons possess the helicity $s$ provided that
\begin{equation}\label{circ_pol}
    J'_m(\vs mz)=-s\vs(n_k-1/z)J_m(\vs mz),\qquad J_1(\bar{k}_0n_k)=s\vs n_kJ_0(\bar{k}_0n_k).
\end{equation}
The linear polarization appears when $A(1)=A(-1)=1/2$, i.e.,
\begin{equation}\label{linear_pol}
    (n_k-1/z)J_m(\vs mz)J'_m(\vs mz)=0,\qquad J_0(\bar{k}_0n_k)J_1(\bar{k}_0n_k)=0.
\end{equation}
The second equalities in \eqref{circ_pol} and \eqref{linear_pol} correspond to the case $m=0$. For $m\neq0$, the equations \eqref{circ_pol} and \eqref{linear_pol} should be solved with account for the radiation spectrum \eqref{spectrum} written in the form of the last equality in \eqref{polariz}. For $m=0$, the condition of the existence of VC radiation, $\bar{\chi}\geqslant\bar{\chi}_c$, should be satisfied.

If one regards the degree of polarization \eqref{polariz} as a function of $n_k$, then the regions where the radiation of photons with a definite helicity dominates are separated by the roots of the equation \eqref{linear_pol} and the regions with the different signs of the helicity are interlaced. Developing \eqref{polariz} for $m\neq0$ as a series in $n_k$, it is not difficult to see that for $n_k\rightarrow0$, i.e., in the ultraparaxial regime, the radiation with $s\sgn(m)=1$ prevails. Furthermore, the positivity of the photon energy \eqref{spectrum} implies that, for $\bar{\chi}<\bar{\chi}_c$, the photons are radiated with $\vs\sgn m=1$.

For $m\neq0$, the graphical solution of the first equation in \eqref{linear_pol} is given in Fig. \ref{znk_plots}. We describe it below. Let $j_{m,p}$ and $j'_{m,p}$, $p=\overline{1,\infty}$, be the positive roots of $J_m(x)$ and $J'_m(x)$, respectively, and
\begin{equation}
    b_{m,p}:=j_{m,p}/|m|,\qquad c_{m,p}:=j'_{m,p}/|m|.
\end{equation}
The properties of zeros of the Bessel functions (see, e.g., \cite{NIST}) imply that $b_{m,p}>1$, $c_{m,p}>1$ and $b_{m,p}\rightarrow1$, $c_{m,p}\rightarrow1$ for $|m|\rightarrow\infty$. For $\bar{\chi}<\bar{\chi}_c$, the function $z(n_k)$ has a unique maximum at the point
\begin{equation}\label{nk0}
    n_k=n_k^0=\sqrt{\bar{\chi}_c-\bar{\chi}},\qquad z(n_k^0)=1/n_k^0,
\end{equation}
and the inflection point at $n_k=\sqrt{3}n_k^0$. At the maximum, $z(n^0_k)\geqslant1$ when $\bar{\chi}\in[K^{-2},\bar{\chi}_c)$. The plots of $z(n_k)$ for $\bar{\chi}\geqslant\chi_c$ are given in Fig. \ref{znk_plots}. As is seen from Fig. \ref{znk_plots}, the properties of polarization of radiation for $m\neq0$ are as follows.

\begin{figure}[tp]
\centering
\includegraphics*[width=0.473\linewidth]{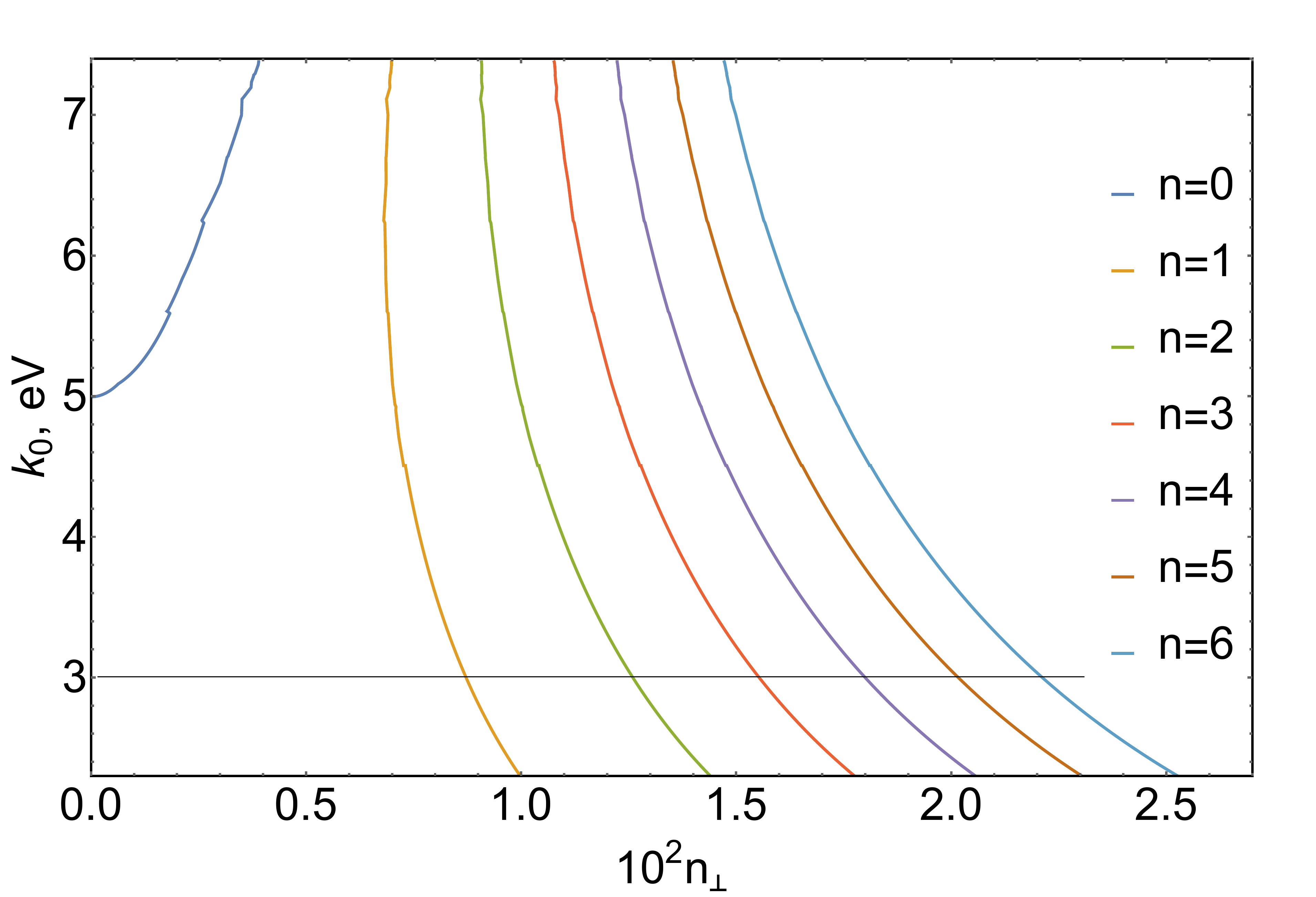}\;
\includegraphics*[width=0.48\linewidth]{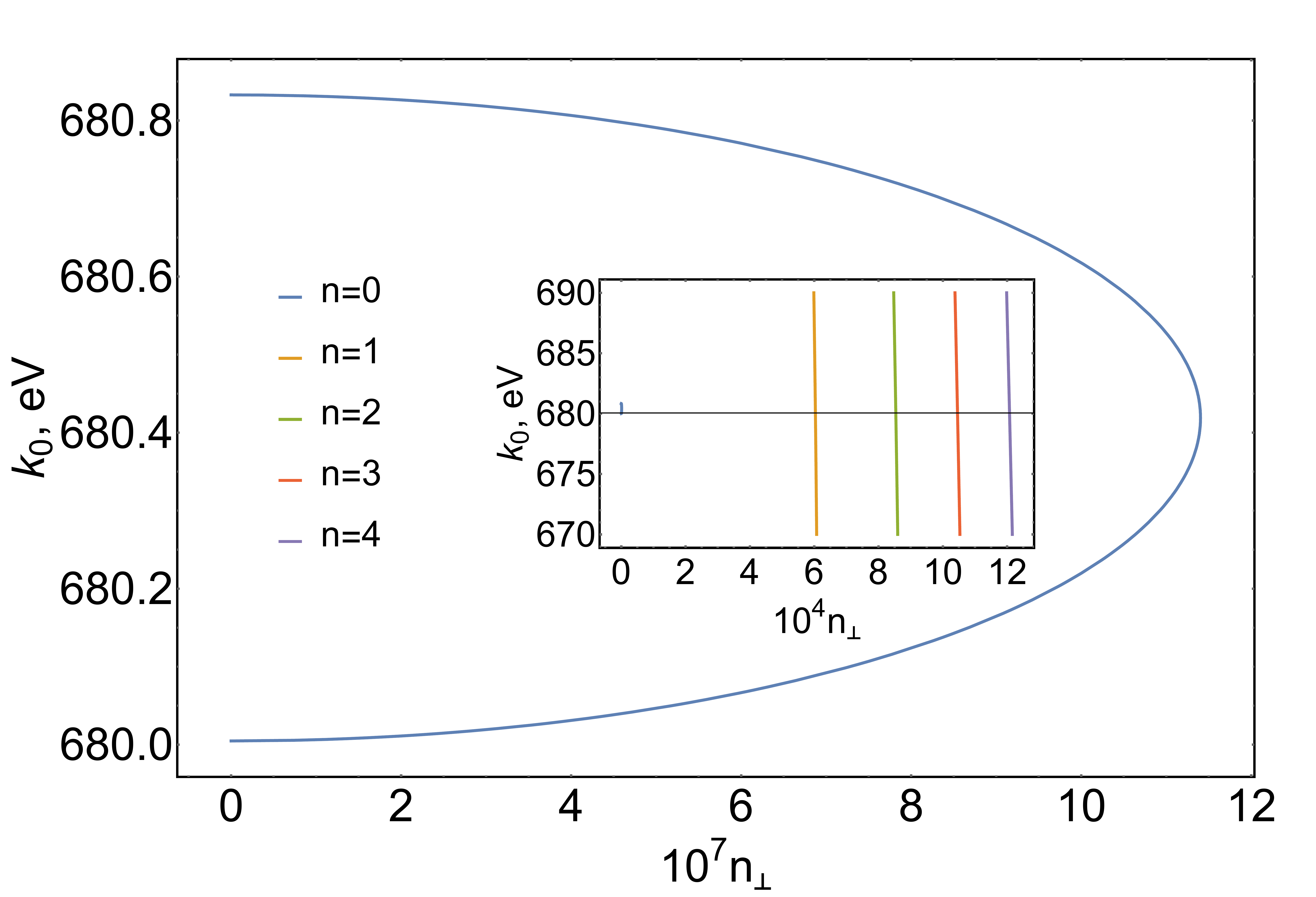}
\caption{{\footnotesize On the left panel: The dependence $k_0(n_\perp)$ for the harmonics $n=\overline{0,6}$ of radiation of twisted photons in the undulator filled with helium under the pressure $4$ atm and the temperature $0$ ${}^o$C. The experimental data for permittivity are taken from \cite{RefrIndex}. On the right panel: The dependence $k_0(n_\perp)$ for the harmonics $n=\overline{0,4}$ of radiation of twisted photons in the undulator filled with xenon under the pressure $1/2000$ atm and the temperature $0$ ${}^o$C. The data for permittivity near the $M$-edge of xenon are taken from \cite{BazylZhev,BazylevXrayVC}. For simplicity, the peak is approximated by a Guassian with the center energy $k_0=680$ eV and the dispersion $15$ eV. This Gaussian is added to the plasma permittivity \eqref{plasm_perm} and the maximum of the obtained permittivity is fitted to the data \cite{BazylZhev,BazylevXrayVC}. One should bear in mind that in \cite{BazylZhev,BazylevXrayVC} the data are given for the pressure $0.2$ atm. The thin horizontal line is $k_0=680$ eV. This energy is slightly below the formation threshold of VC radiation.}}
\label{spectra_plots}
\end{figure}

For $\bar{\chi}\leqslant K^{-2}$ there is a single solution to the first equation in \eqref{linear_pol} at $n_k=n^0_k$. Therefore, for  $n_k<n_k^0$, the radiation with $s\vs=1$ dominates, whereas, for $n_k>n_k^0$, the main contribution comes from the radiation with $s\vs=-1$. We will refer to the parameter space, where $s\vs=-1$ for the main contribution to radiation, as the domain with inverted radiation polarization. In the absence of the medium, the existence of this domain can easily be explained without any calculations. To this end, one needs to find the helicity of created radiation in the reference frame where the charge is at rest on average (the synchrotron frame). In this frame, the radiation with $s\vs=1$ dominates in the half-space $z>0$, whereas the radiation with $s\vs=-1$ prevails in the half-space $z<0$. Then, by using the Lorentz transformation, one passes to the laboratory frame and takes into account that the photon helicity is Lorentz-invariant (see for more detail \cite{Bord.1,BKL4}). The value $n_k=n_k^0$ with $\bar{\chi}=0$ corresponds to the angle at which the orbit plane of a charge in the synchrotron frame is seen in the laboratory frame. Formula \eqref{nk0} gives the value of this angle with account for the nonvanishing electric susceptibility, $\bar{\chi}\neq0$. For $\bar{\chi}=K^{-2}$, we have $n_k^0=1$ (see Fig. \ref{znk_plots}).

If $\bar{\chi}\in[\bar{\chi}_c, K^{-2})$, then $n_k\geqslant n_k^{(+)}$ corresponds to the domain with inverted radiation polarization for any harmonic, and $n_k\leqslant n_k^{(-)}$ is the region with the usual radiation polarization (the radiation with $s\vs=1$ dominates) for any number of the radiation harmonic. Here
\begin{equation}
    n^{(\pm)}_k:=1\pm\sqrt{1+\bar{\chi}-\bar{\chi}_c},
\end{equation}
which have been found from the condition $z(n_k^{(\pm)})=1$. For $n_k\in(n_k^{(-)},n_k^{(+)})$, the regions with the different signs of $s\vs$ are interlaced and separated by the roots of the first equation in \eqref{linear_pol}, viz., by the solutions of the equations
\begin{equation}
    z=b_{m,p}, \qquad z=c_{m,p}, \qquad n_k=n_k^0,
\end{equation}
where $p=\overline{1,\infty}$. The energy of radiated photons is found from the intersection point of the plots $z=z(n_k)$ and $z=\bar{k}_0n_k/|m|$. Notice that the dipole approximation is not applicable in the domain $|z|\gtrsim1$.

For $\bar{\chi}>\bar{\chi}_c$, there appears the region, $n_k<|n_k^0|$, where the radiation with the anomalous Doppler effect is observed. In this region, $\vs\sgn m=-1$. The domain with inverted radiation polarization for any radiation harmonic is specified by the inequalities $n_k\leqslant |n_k^{(-)}|$ or $n_k\geqslant n_k^{(+)}$. For $n_k\in(|n_k^{(-)}|,n_k^{(+)})$, the regions with the different signs of $s\vs$ are interlaced and separated by the roots of the equations
\begin{equation}
    |z|=b_{m,p}, \qquad |z|=c_{m,p},
\end{equation}
where $p=\overline{1,\infty}$. The energy of a radiated photon is found from the last equality in \eqref{polariz}.

When $m=0$ and $\bar{\chi}\geqslant\bar{\chi}_c$, i.e., the VC radiation is considered, it is convenient to introduce the variable $y:=\bar{k}_0n_k$, where $n_k=|n_k^0|=\sqrt{\bar{\chi}-\bar{\chi}_c}$, for the analysis of the degree of radiation polarization. If $y<j_{0,1}$ then the radiation with $s\vs=1$ dominates. In increasing $y$, the intervals where the radiation with $s\vs=1$ or $s\vs=-1$ dominates are interlaced as
\begin{equation}
    s\vs=
    \left\{
      \begin{array}{ll}
        1, & \hbox{$y\in(j_{1,p},j_{0,p+1})$;} \\
        -1, & \hbox{$y\in(j_{0,p},j_{1,p})$,}
      \end{array}
    \right.\quad p=\overline{1,\infty}.
\end{equation}
The VC radiation is completely circularly polarized and consists of photons with the helicity $s$ provided that
\begin{equation}
    \bar{k}_0=s\vs\frac{yJ_0(y)}{J_1(y)}.
\end{equation}
In this case, the orbital angular momentum of radiated photons becomes
\begin{equation}
    l=m-s=-s.
\end{equation}
In particular, in the dipole regime, $y\ll1$, we have
\begin{equation}\label{VC_twist}
    \bar{k}_0=2,\qquad s\vs=1,\qquad 2|n_k^0|\ll1.
\end{equation}
Notice that for this photon energy the radiation of harmonics with $n=\vs m\geqslant1$ is not described by the formulas of the dipole approximation, because in this case
\begin{equation}
    \bar{k}_0n_k=2\sqrt{n+\bar{\chi}-\bar{\chi}_c}>2.
\end{equation}
Therefore, the radiation at these harmonics is not suppressed and its intensity can exceed the intensity of the VC radiation (see Figs. \ref{VC_pol_plots} and \ref{VC_pol_beam_plots}).

\begin{figure}[tp]
\centering
\includegraphics*[width=0.48\linewidth]{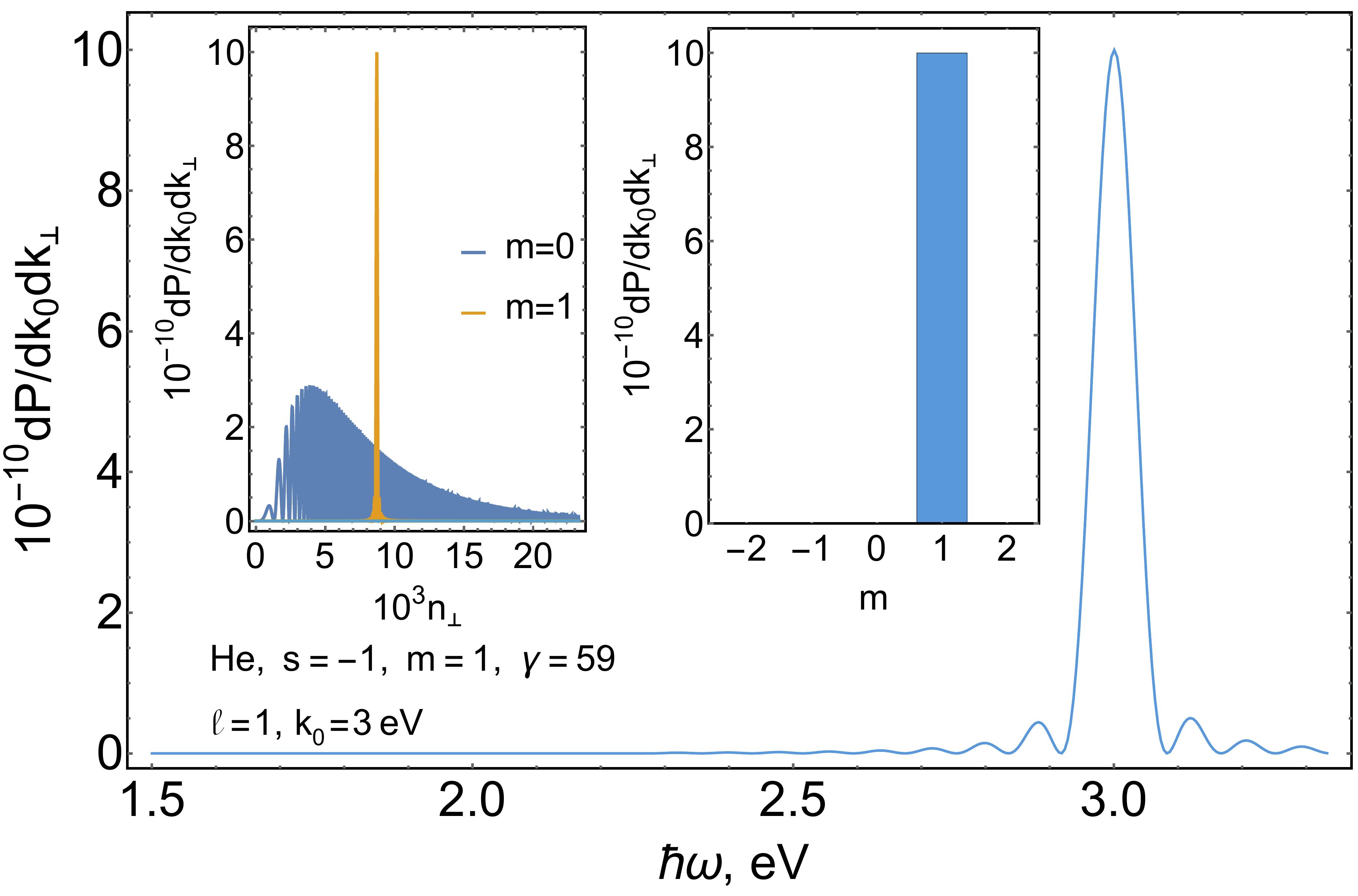}\;
\includegraphics*[width=0.473\linewidth]{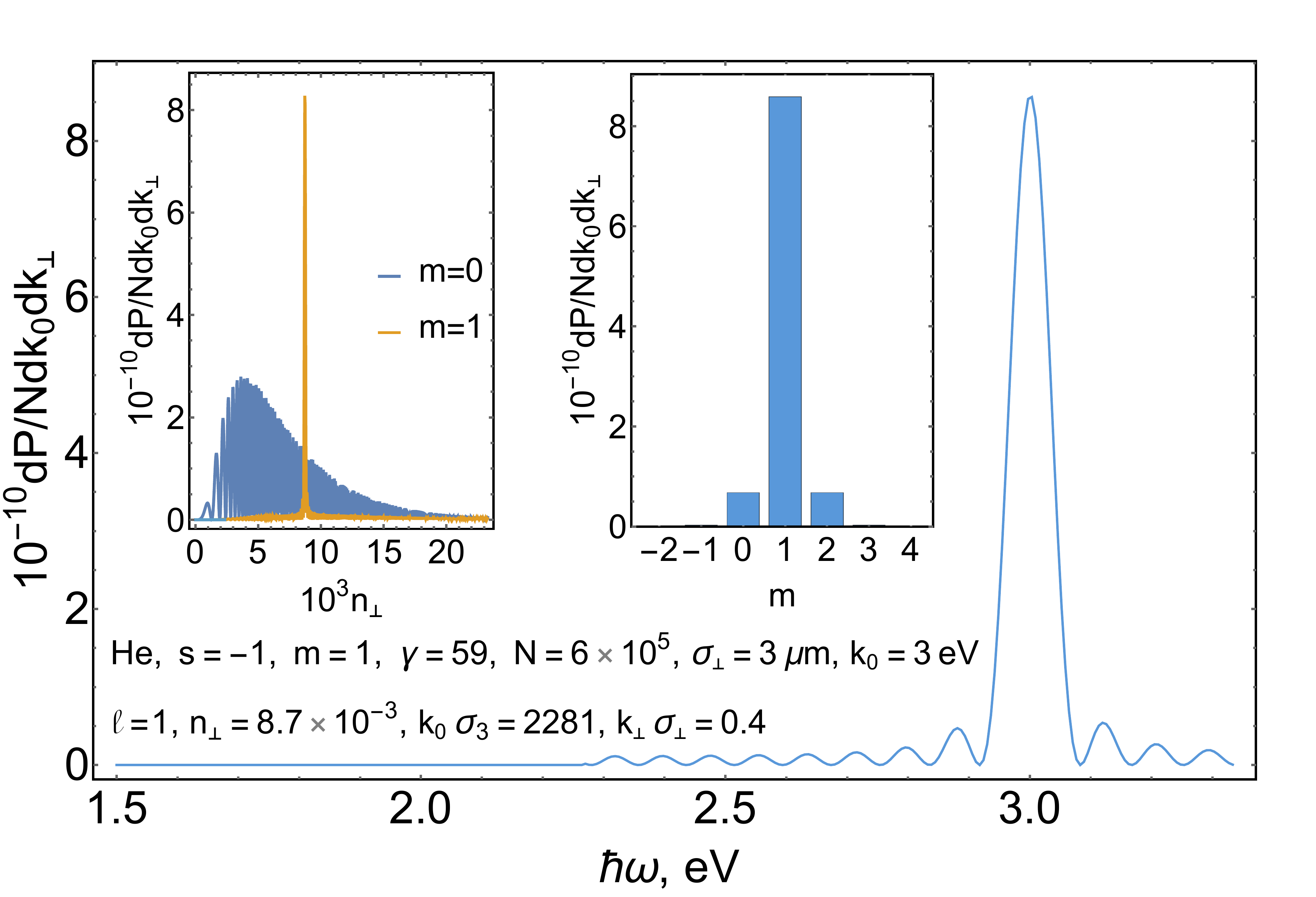}
\caption{{\footnotesize On the left panel: The average number of twisted photons produced by one proton moving in the helical undulator with the chirality $\vs=1$. The Lorentz factor is $\ga=59$, $E=55.4$ GeV. The number of undulator periods $N_u=40$ and the period $\la_0=1$ cm. The undulator strength parameter is $K=8.7\times10^{-3}$, which corresponds to the magnetic field strength $H=241.5$ kG in it. The energy spectrum of photons is given on the left panel in Fig. \ref{spectra_plots}. The ratios $n_\perp^2/q\approx1.1\times 10^6$ and $K^2/(q\ga^2)\approx462$ that means that the multiple scattering can be neglected. The radiation at the first harmonic with $s=-1$ dominates and so the radiation of photons with projection of the orbital angular momentum $l=2$ prevails. On the right panel: The same as on the left panel but for the Gaussian beam of protons. The coherent contribution to radiation is strongly suppressed.}}
\label{protons_plots}
\end{figure}

Now we consider the orbital angular momentum of twisted photons radiated at the harmonics with $n\neq0$. At these harmonics, the projection of the orbital angular momentum of photons is
\begin{equation}
    l=m-s=\vs(n-s\vs).
\end{equation}
When $n\geqslant1$ and $s\vs=1$, this relation reproduces the known selection rule for the orbital angular momentum of twisted photons radiated by a helical undulator \cite{HemMar12,KatohSRexp,KatohPRL,BHKMSS,SasMcNu,RibGauNin14}. However, at the same harmonic but in the domain with inverted radiation polarization, the absolute value of the orbital angular momentum of a radiated photon is by $2\hbar$ more than in the region with the usual polarization. The same property has the radiation in the parameter space where the anomalous Doppler effect is realized, $n\leqslant-1$, but for $s\vs=1$. In particular, selecting suitably the energy of the observed photons, the parameters of the radiating charged particle and the medium, one can secure that the first harmonic of the undulator radiation consists of the twisted photons with $l=2\vs$.

As for the undulator radiation in a vacuum, this effect was discussed in \cite{BKL4}. As is seen from the analysis given above, in a vacuum, $\bar{\chi}=0$, and for the photon energy
\begin{equation}\label{inv_polar}
    \bar{k}_0<(n_k^0)^{-2}=(1+K^{-2})^{-1}\;\Leftrightarrow\;k_0<\frac{\omega\ga^2}{1+K^2},\qquad n^2_\perp\ga^2=\frac{2\omega\ga^2}{k_0}-1-K^2,
\end{equation}
the photons at the first harmonic possess the projection of the orbital angular momentum $l=2\vs$. The presence of the medium with $\chi>0$ allows one to decrease the cone opening down to $n^0_k=1$ and, as follows from \eqref{spectrum} and \eqref{inv_polar}, to increase the energy of photons possessing $l=2\vs$ at the first harmonic.

\begin{figure}[tp]
\centering
\includegraphics*[width=0.48\linewidth]{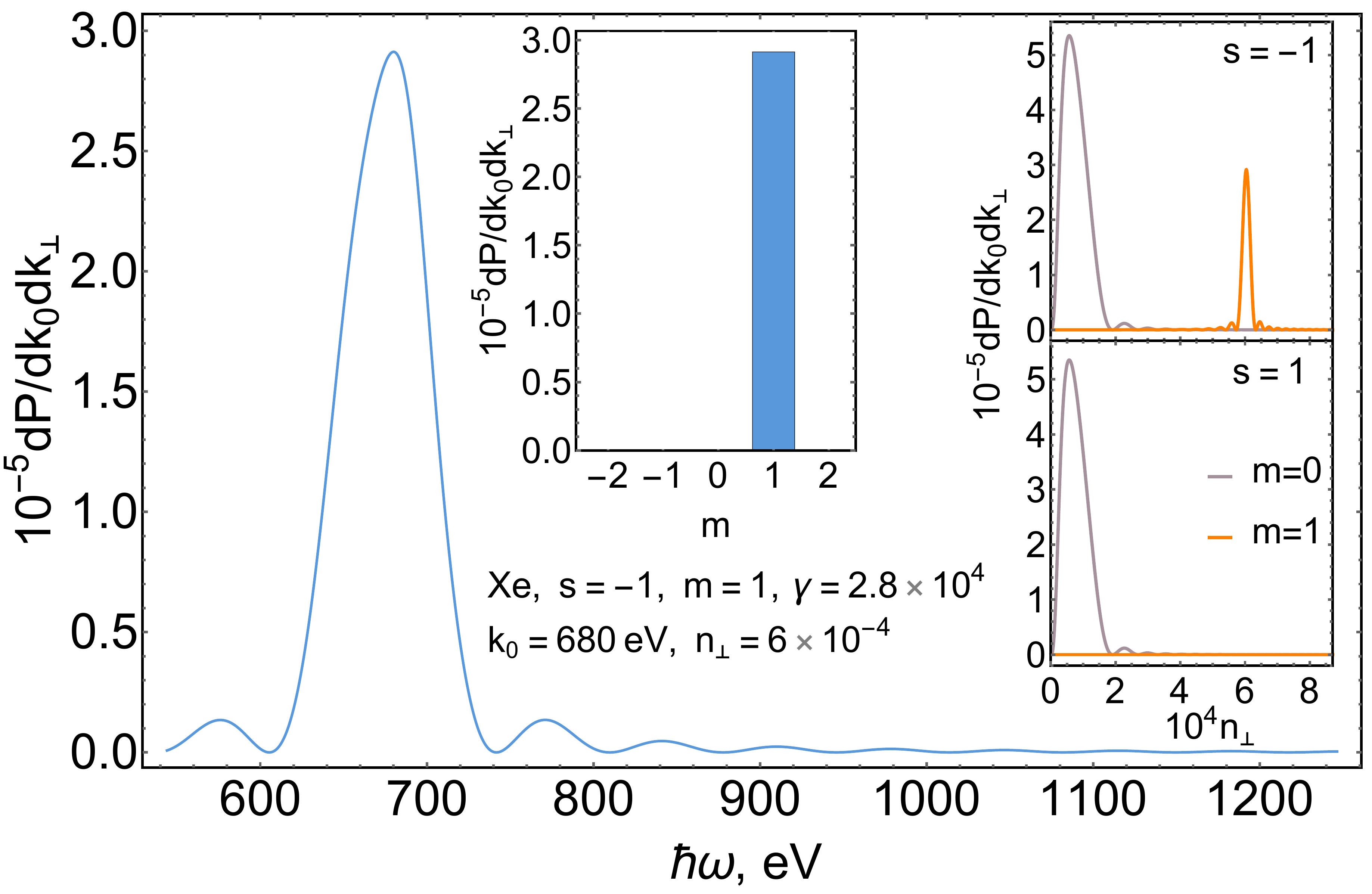}\;
\includegraphics*[width=0.48\linewidth]{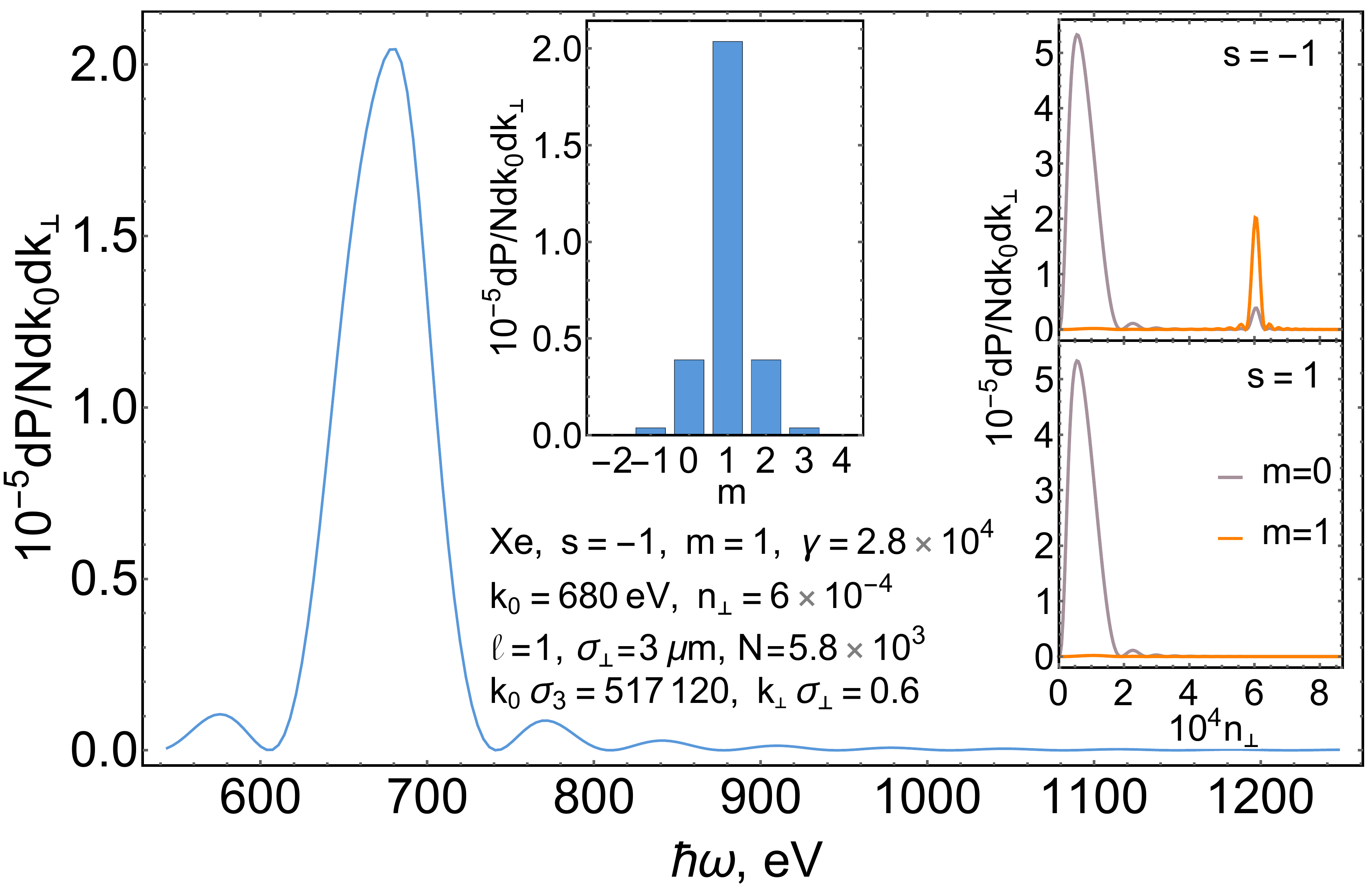}
\caption{{\footnotesize On the left panel: The average number of twisted photons produced by one electron moving in the helical undulator with the chirality $\vs=1$. The Lorentz factor is $\ga=2.84\times 10^4$, $E=14.5$ GeV. The number of undulator periods $N_u=10$ and the period $\la_0=1$ cm. The undulator strength parameter $K=1/10$, which corresponds to the magnetic field strength $H=1.5$ kG in it. The energy spectrum of photons is given on the right panel in Fig. \ref{spectra_plots}. The ratios $n_\perp^2/q\approx 6.9\times 10^6$ and $K^2/(q\ga^2)\approx1.7$ that means that the multiple scattering can be neglected. The photons at the first harmonic $k_0=680$ eV possess $s=-1$ and so they have the projection of the orbital angular momentum $l=2$. The large peak with $m=0$ near the origin is the contribution of transition radiation. On the right panel: The same as on the left panel but for the Gaussian beam of electrons. The coherent contribution to radiation is strongly suppressed.}}
\label{Xe_Xray_plots}
\end{figure}

\paragraph{Dipole regime.}

As a particular case of the above general relations, we consider the polarization of radiation with $m\neq0$ in the dipole regime when $|mz|\ll1$. Then formula \eqref{circ_pol} implies that the undulator radiation is completely circularly polarized and consists of the photons with the helicity $s$ provided that
\begin{equation}
    n_k=
    \left\{
      \begin{array}{ll}
        \frac{|m|z}{2(|m|+1)}, & \hbox{$s\sgn(m)=1$, $s\vs=1$;} \\[0.3em]
        \frac2{z}, & \hbox{$s\sgn(m)=-1$, $s\vs=-1$.}
      \end{array}
    \right.
\end{equation}
Whence, in the first case,
\begin{equation}
    \bar{k}_0=2(|m|+1),\qquad n_k^2=\frac{|m|}{|m|+1}+\bar{\chi}-\bar{\chi}_c,\qquad \bar{k}_0n_k\ll1.
\end{equation}
And, as it was established above in general, the radiation with $s\sgn(m)=1$ dominates for $n_k<|n_k^{(-)}|$. In that case, the orbital angular momentum of the photons radiated at the first harmonic is $l=0$. In the second case,
\begin{equation}
    \bar{k}_0=2|m|/n_k^2,\qquad\bar{\chi}=\bar{\chi}_c,\qquad 2|m|/n_k\ll1.
\end{equation}
For these parameters, the radiation completely consists of the twisted photons with helicity $s=-\vs$ and projection of the orbital angular momentum $l=2\vs$ (see Figs. \ref{spectra_plots}, \ref{protons_plots}, and \ref{Xe_Xray_plots}). This value of the orbital angular momentum can be shifted by an integer number employing the coherent radiation of the helically microbunched beams of particles in undulators \cite{BKLb,HKDXMHR}, the center of such a beam moving along the trajectory \eqref{traj_undul}.

\subsection{Planar wiggler}\label{Plan_Wigg_Sec}

The trajectory of a charged particle propagating in the planar undulator has the form \eqref{traj_undul} with (see for details, e.g., \cite{Bord.1})
\begin{equation}
    r_{\pm}=\pm\sqrt{2}i\frac{\beta_{3} K}{\omega \gamma}\sin(\omega t), \qquad r_{3}=-\frac{\beta_{3} K^{2}}{4\omega \gamma^{2}}\sin(2\omega t).
\end{equation}
Let us find the average number of twisted photons radiated by the charged particle taking into account that $K/\gamma\ll1$. Then the velocity components of the charge moving in the medium become
\begin{equation}
    \dot{x}_{\pm}\approx \pm \sqrt{2}i\frac{K}{\gamma}\cos(\omega t), \qquad \dot{x}_{3}\approx1 .
\end{equation}
The integrals \eqref{I_integrals} are evaluated in the same way as for the planar wiggler in a vacuum studied in Sec. 5.B.2 of \cite{BKL2}
\begin{equation}
    I_{3}=\sum_{n=-\infty}^{\infty} \vf_{n}f_{n,m}, \qquad I_{\pm}=\mp\frac{s'\e^{1/2}\pm n'_{3}}{n_{k}}\sum_{n=-\infty}^{\infty} \vf_{n} f_{n,m}^{\pm},
\end{equation}
As before, the functions $\vf_{n}$ are defined by formula \eqref{vfn} and, for brevity, the following notation has been introduced
\begin{equation}\label{f}
\begin{split}
    f_{n,m}&:=\pi (1+ (-1)^{n+m})\sum_{k=-\infty}^{\infty} J_{k}\Big(\frac{\beta_{3} k'_{3} K^{2} }{4 \omega \gamma^{2}}\Big) J_{(n-m)/2+k}\Big(\frac{\beta_{3} K k_{\bot}}{\sqrt{2} \omega \gamma}\Big) J_{(n+m)/2+k}\Big(\frac{\beta_{3} K k_{\bot}}{\sqrt{2} \omega \gamma}\Big), \\
    f_{n,m}^{\pm}&:= \frac{f_{n+1,m\mp1} +f_{n-1,m\mp1}}{\sqrt{2}}.
\end{split}
\end{equation}
Neglecting the contribution of the transition radiation, the average number of twisted photons radiated by one particle takes the form
\begin{multline}\label{planar}
    dP(s,m,k_\perp,k_3)=\bigg|zea\sum_{n=-\infty}^{\infty}\vf_{n}\Big[\Big(\frac{1}{\e}+\frac{n_{3}}{n'_{3}}\Big)\Big(f_{n,m}-\frac{n'_{3}}{2 n_{k}}\big(f_{n,m}^{+}+f_{n,m}^{-}\big)\Big)+\frac{s}{2 n_{k}}\Big(1+\frac{n_{3}}{n'_{3}}\Big)\big(f_{n,m}^{-}-f_{n,m}^{+}\big)\Big]\\
    +(k'_{3}\leftrightarrow - k'_{3})\bigg|^2 n_\perp^3\frac{dk_\perp dk_3}{64\pi^2}.
\end{multline}
The radiation spectrum is given by formula \eqref{harmonics} and the analysis of its peculiarities is presented in Sec. \ref{spectrum_sect}. Taking into account that
\begin{equation}
    f^+_{n,m}=f^-_{n,-m},
\end{equation}
it is easy to see that \eqref{planar} obeys the reflection symmetry \eqref{refl_symm}.

\begin{figure}[tp]
\centering
\includegraphics*[width=0.48\linewidth]{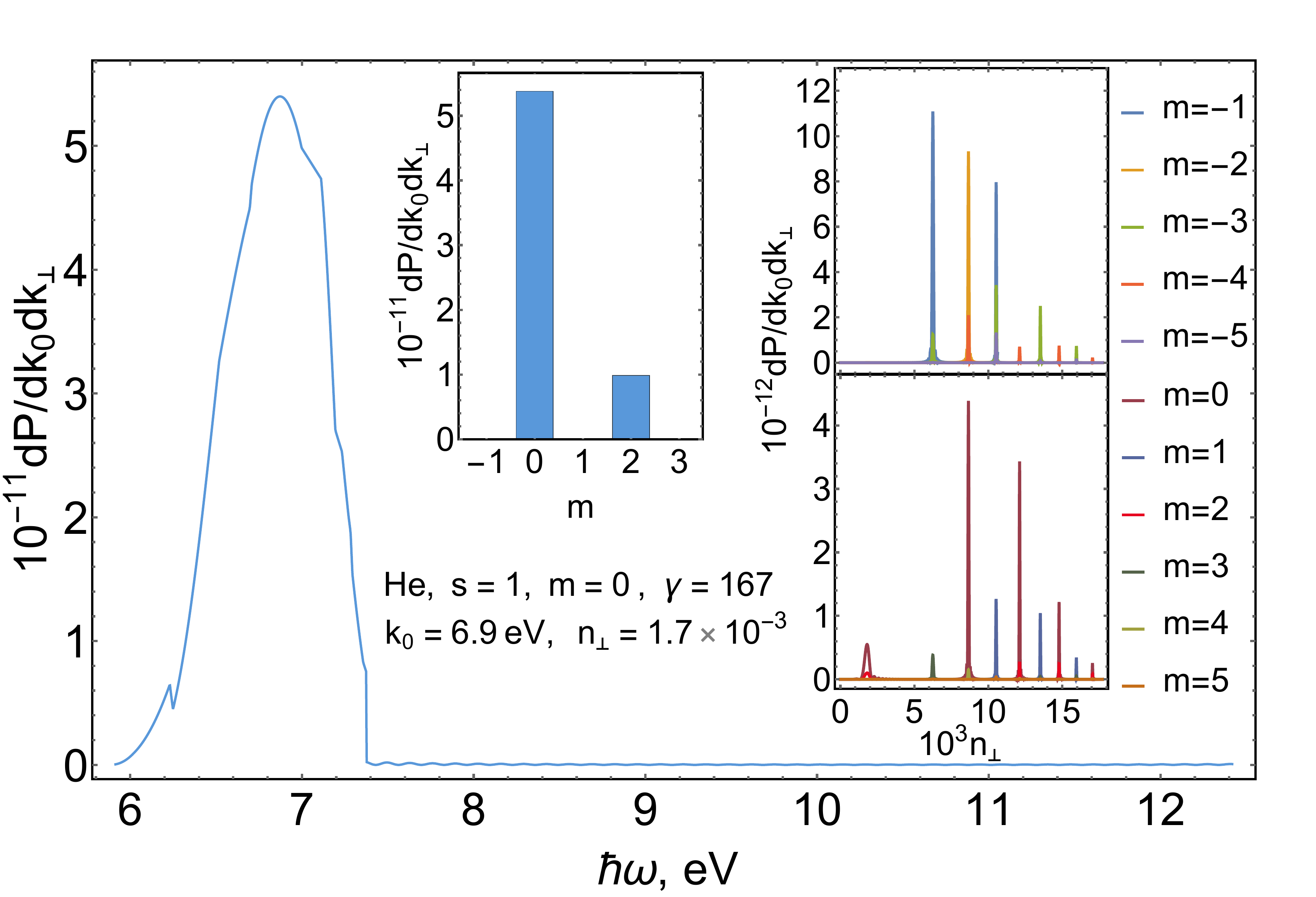}\;
\includegraphics*[width=0.48\linewidth]{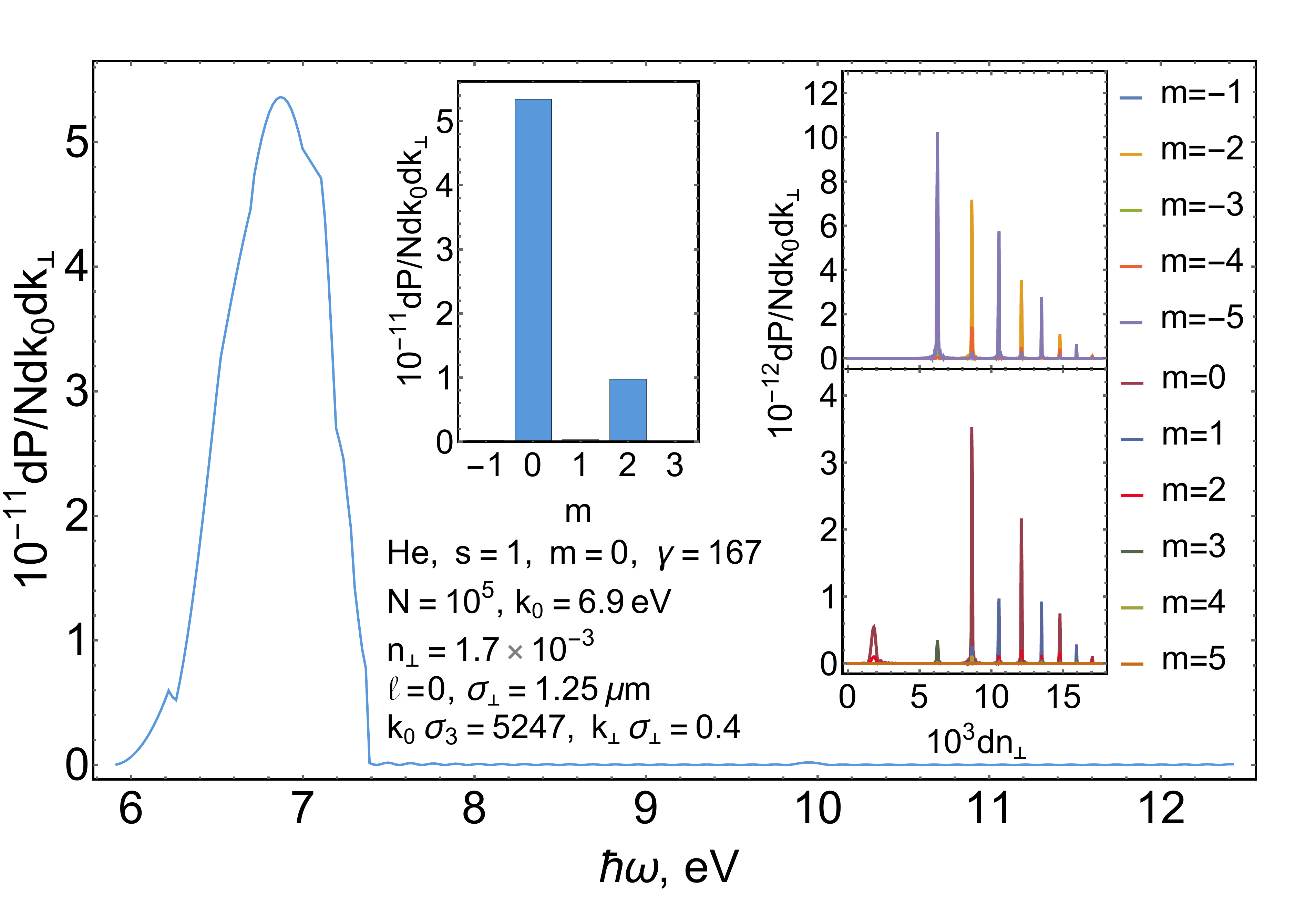}
\caption{{\footnotesize The average number of twisted photons produced by electrons in the planar wiggler. The parameters of the wiggler and the energy of electrons are the same as in Fig. \ref{VC_pol_plots}. On the left panel: The average number of twisted photons produced by one electron moving in such planar wiggler. The energy of photons at the zeroth harmonic (the VC radiation) is taken as $k_0=6.9$ eV. The ratios $n_\perp^2/q\approx1.0$ and $K^2/(q\ga^2)\approx12$ that means that the multiple scattering can be neglected. We see that the selection rule $m+n$ is an even number is satisfied. The reflection symmetry \eqref{refl_symm} also holds. At the zeroth harmonic, the radiation of photons with $s=-1$ is suppressed. Therefore the radiation of photons with projection of the orbital angular momentum $l=1$ dominates and there is a small admixture of photons with $l=3$. On the right panel: The same as on the left panel but for the Gaussian beam.}}
\label{Planar_wig_plots}
\end{figure}

If the thickness of the dielectric medium $L$ is large, the contribution of the reflected wave is suppressed in comparison with the contribution of the direct wave. Moreover, the contribution of the terms describing the interference between different harmonics is small in comparison with the values of \eqref{planar} at the peaks of the harmonics. Having neglected these small contributions, we deduce the expression for the average number of twisted photons radiated at the $n$-th harmonic
\begin{equation}\label{planarn}
    dP(s,m,k_\perp,k_3)=|zea\vf_{n}|^{2}\bigg|\Big(\frac{1}{\e}+\frac{n_{3}}{n'_{3}}\Big)\Big(f_{n,m}-\frac{n'_{3}}{2 n_{k}}\big(f_{n,m}^{+}+f_{n,m}^{-}\big)\Big)+\frac{s}{2 n_{k}}\Big(1+\frac{n_{3}}{n'_{3}}\Big)\big(f_{n,m}^{-}-f_{n,m}^{+}\big)\bigg|^2 n_\perp^3\frac{dk_\perp dk_3}{64\pi^2}.
\end{equation}
As follows from the explicit expressions for the functions $f_{n,m}, f_{n,m}^{\pm}$ entering into \eqref{planar} and \eqref{planarn}, the number $n+m$ is an even one for radiated twisted photons. This selection rule was obtained in \cite{BKL2} for the radiation of twisted photons by a planar undulator in a vacuum and it was shown in \cite{BKL6} that this selection rule is preserved when the quantum recoil is taken into account. In particular, this selection rule implies that the VC radiation in the planar undulator corresponding to $n=0$ consists of the twisted photons with an even projection of the total angular momentum $m$ and not just with $m=0$. The plots of the average number of twisted photons radiated by electrons in the planar wiggler filled with helium are presented in Fig. \ref{Planar_wig_plots}.


\section{Conclusion}

Let us sum up the results. We described the properties of radiation of twisted photons produced by undulators filled with a homogeneous dielectric medium. Both the dipole and wiggler regimes were considered. We started with the general formula for the probability to detect a twisted photon radiated by a charged particle passing through a dielectric plate derived in \cite{BKL5}. We proved that the selection rules established in \cite{BKL3} for radiation of twisted photons by charged particles in a vacuum also hold in the presence of a dielectric plate. In particular, we proved the reflection symmetry property \eqref{refl_symm} for the radiation probability of twisted photons by planar currents.

Then the formulas for the average number of twisted photons radiated by an undulator filled with a dielectric medium were deduced. We studied in detail the undulator radiation in the dipole regime and the ideal helical and planar wigglers. We analyzed the spectrum of energies of radiated photons paying a special attention to the case of a plasma permittivity. In particular, we showed that, for sufficiently large plasma frequencies \eqref{prohib_harm}, the lower harmonics of undulator radiation do not form (see Figs. \ref{plsm_perm_plots} and \ref{znk_plots}).

We also investigated the spectrum of twisted photon radiation with respect to the projection of the total angular momentum $m$ and the orbital angular momentum $l$. We showed that, in the general dipole case, the undulator radiation mainly consists of the twisted photons with $m=\{-1,0,1\}$ provided the lower harmonics are not prohibited by the energy spectrum at a given energy. Recall that the undulator dipole radiation in a vacuum is mainly comprised by the twisted photons with $m=\pm1$ \cite{BKL2}. In the case of the undulator dipole radiation filled with a medium, the radiation with $m=0$ corresponds to the VC radiation. In the ultraparaxial approximation \eqref{ultraparaxial}, we found that the most part of twisted photons of the dipole undulator radiation with $m=\pm1$ possess the orbital angular momentum $l=0$.

In considering a helical wiggler filled with a dielectric medium, we found that the selection rule $m=\vs n$, where $n$ is the harmonic number and $\vs=\pm1$ is the chirality of the helical trajectory, is satisfied. This selection rule is the same as in the vacuum case but the harmonic number can be negative for the medium with the electric susceptibility $\chi>0$ due to the anomalous Doppler effect. The case $n=0$ corresponds to the VC radiation. A peculiar polarization properties of radiation created by the helical wiggler filled with a medium allows one to produce the radiation with a well-defined orbital angular momentum $l$ in the paraxial regime (see Figs. \ref{VC_pol_plots}, \ref{VC_pol_beam_plots}, \ref{protons_plots}, and \ref{Xe_Xray_plots}). We described thoroughly these polarization properties and found the parameter space where the radiation with a given $l$ dominates.

For example, for any given harmonic, we found the domains with inverted polarization where the radiation with $s\vs=-1$ prevails, $s$ being the helicity of radiated twisted photons. Such domains exist already for the vacuum undulator radiation (see, e.g., \cite{Bord.1,BKL4}). The presence of a dielectric medium allows one to increase the radiation yield and the energy of photons in these domains. In the paraxial regime, $l=\vs(n-s\vs)$ and so the absolute value of the orbital angular momentum at a given harmonic, $n\geqslant1$, is by $2\hbar$ more in the domain with inverted polarization than in the region where the usual polarization, $s\vs=1$, prevails. As for the harmonics $n\leqslant-1$, where the anomalous Doppler effect appears, the situation is reverse. For a given harmonic, the orbital angular momentum is by $2\hbar$ more in the domain with $s\vs=1$ than in the domain with inverted polarization.

Besides, we found the parameter space where the VC radiation is almost completely circularly polarized. As long as $m=0$ for the VC radiation produced in the helical undulator, the photons of VC radiation possess a definite nonzero orbital angular momentum in this case, provided the paraxial approximation is valid. For example, the VC radiation is constituted by the twisted photons with orbital angular momentum $l=-\vs$ at the photon energy $k_0\approx2\omega\ga^2/K^2$ near the threshold of the VC radiation \eqref{VC_twist} (see Figs. \ref{VC_pol_plots} and \ref{VC_pol_beam_plots}).

The spectrum over $m$ of the twisted photons produced in the planar wiggler was also studied. We proved that the selection rule, $m+n$ is an even number, is fulfilled for this radiation. It has the same form as for the planar wiggler radiation in a vacuum \cite{BKL2,BKL6}. It was also found that the VC radiation produced by charged particles in such a wiggler consists of twisted photons with even projections of the total angular momentum and not just with $m=0$ (see Fig. \ref{Planar_wig_plots}). Of course, the reflection symmetry of the probability of radiation of twisted photons \eqref{refl_symm} holds in this case.

All the above mentioned properties are valid for the undulator radiation produced by one charged particle. We investigated how these properties change when the radiation of a beam of particles is considered (see Figs. \ref{VC_pol_beam_plots}, \ref{protons_plots},  \ref{Xe_Xray_plots}, and \ref{Planar_wig_plots}). As expected, the properties of radiation of twisted photons by a structureless beam of charged particles are the same as for the radiation generated by one particle when $k_\perp\s_\perp\lesssim1$, where $\s_\perp$ is the transverse size of the beam \cite{BKb,BKLb}. The use of periodically microbunched beams of particles allows one to amplify the intensity of radiation by means of the coherent effects. If, in addition, such a microbunched beam is helical, the total angular momentum of radiated twisted photons can also be increased (see Fig. \ref{VC_pol_beam_plots}). Nowadays, the techniques are elaborated allowing one to produce such a coherent radiation up to X-ray spectral range \cite{Gover19rmp,PRRibic19,Rubic17,Hemsing7516,HemStuXiZh14}.

As examples, we described the radiation of twisted photons produced by electron and proton beams in the undulators filled with helium in the ultraviolet and X-ray spectral ranges. Moreover, we considered the production of X-ray twisted photons with $l=2$ by the electron beam propagating in the undulator filled with xenon. This radiation is created near the photoabsorption $M$-edge of xenon.

Thus we see that the presence of a dispersive medium inside of the undulator offers additional possibilities for generation of hard twisted photons with desired properties. We did not study in this paper the effect of a periodic modulation of the dielectric medium on the properties of radiation of twisted photons \cite{Appolonov,TMikaelian,Ginzburg,BazylZhev,BKL5}. This will be the subject of our future research. Besides, as was shown in \cite{FedSmir,BazylZhev}, the VC radiation can be generated in the gamma ray spectral range near the lines of M\"{o}ssbauer transitions. The usual VC radiation is an equiprobable mixture of twisted photons with $l=\pm1$ in the paraxial regime (see, e.g., \cite{Kaminer,OAMeVCh,BKL7,BKL5}). The theory developed in the present paper can be employed to twist these gamma rays and to make them to be in an eigenstate of the orbital angular momentum operator.


\paragraph{Acknowledgments.}

The reported study was funded by RFBR, project number 20-32-70023.

\end{document}